\documentclass[]{aastex62}
\usepackage{amsmath}
\usepackage[english]{babel}
\usepackage[]{graphicx}
\usepackage{multirow}
\usepackage{placeins}
\usepackage[normalem]{ulem}
\usepackage{txfonts}
\usepackage{xspace}

\usepackage{newtxtext,newtxmath}

\usepackage{mleftright}
\mleftright

\newcommand{\ro}{\varrho}
\newcommand{\br}[1]{\left({#1}\right)}
\newcommand{\pr}[1]{\left\{{#1}\right\}}
\renewcommand{\sq}[1]{\left[{#1}\right]}

\newcommand{\reff}[1]{(\ref{#1})}
\newcommand{\eps}{\epsilon}

\renewcommand{\sc}{x^{\star}}
\newcommand{\xc}{x^{\star}}
\newcommand{\yc}{y^{\star}}
\newcommand{\zc}{\epsilon^{\star}}

\newcommand{\scc}{x^{\star\star}}
\newcommand{\ud}[1]{\mathrm{d}{#1}}
\newcommand{\sgn}[1]{\mathrm{sgn}{\left(#1\right)}}
\newcommand{\kappac}{\kappa^{\star}}

\newcommand{\expp}[1]{\mathrm{e}^{#1}}
\newcommand{\ie}{{\it i.e.}\xspace}
\newcommand{\lrg}[1]{\mbox{\large${#1}$}}
\newcommand{\srg}[1]{\mbox{\normalsize${#1}$}}
\newcommand{\ssrg}[1]{\mbox{\scriptsize${#1}$}}
\newcommand{\xmin}{x_{\mathrm{min}}}

\newcommand{\xzero}{x_{0}}
\newcommand{\emin}{\epsilon_{\mathrm{min}}}

\newcommand{\tr}{\mathrm{Tr}}
\newcommand{\locl}{\lambda_{\mathrm{ocl}}}
\newcommand{\xocl}{x_{\mathrm{ocl}}}
\newcommand{\lsfl}{\lambda_{\mathrm{sfl}}}
\newcommand{\xsfl}{x_{\mathrm{sfl}}}
\newcommand{\epsfl}{\epsilon_{\mathrm{sfl}}}

\newcommand{\tcff}[2]{{{(\!#1\!)}_{{}_{\!#2}}}}

\shorttitle{Cylindrically symmetric spiralling accretion}
\shortauthors{Bratek et al.}

\begin{document}

\title{Cylindrically symmetric spiralling accretion\\ in power-law and logarithmic potentials}

\correspondingauthor{{\L}ukasz Bratek}
\email{lukasz.bratek@pk.edu.pl}

\author[0000-0002-0272-8236]{{\L}ukasz Bratek}
\affiliation{Institute of Physics, Cracow University of Technology, ul.~Podchor\c{a}\.{z}ych 1, PL-30084 Krak\'{o}w, Poland}

\author[0000-0003-0026-5650]{Joanna Ja{\l}ocha}
\affiliation{Institute of Physics, Cracow University of Technology, ul.~Podchor\c{a}\.{z}ych 1, PL-30084 Krak\'{o}w, Poland}

\author{Marek Kutschera}
\affiliation{Institute of Physics, Jagiellonian University, PL-30059  Krak\'{o}w, Poland}

%--------------------------- ABSTRACT ----------------------------------
\begin{abstract}

We study cylindrically symmetric steady state accretion of polytropic test matter spiralling onto the symmetry axis in power-law and logarithmic potentials. The model allows to qualitatively understand the accretion process in a symmetry different from that of the classical Bondi accretion. 
We study the integral curves as level lines of some Hamiltonian and apply this method also to Bondi accretion. The isothermal solutions in power-law potentials (as well as in any radius-dependent potential) can be expressed in exact form in terms of the Lambert $W$ function, while in the case of logarithmic potential exact solutions can be found for any polytropic exponent.

\end{abstract}


%---------------------------- MAIN -------------------------------------

%\defaultscriptratio=0.7
 % \defaultscriptscriptratio=0.5

\section{Introduction}

\noindent
The radial steady flow of polytropic matter  under spherical symmetry is well understood on the basis of Bondi model \citep{1952MNRAS.112..195B}. There is also a generally-relativistic counterpart of the model better suited in close proximity of condensed objects \citep{1972Ap&SS..15..153M}. 
In the context of \citet{parker1958} solar winds,  \citet{cranmer2004} solved the model exactly for the isothermal case in terms of the Lambert $W$ function \citep{wright1959,corless1996}.
Recently, the Bondi model has been considered with other spherical potentials of astrophysical interest. \cite{Jaffe} and \cite{Hernquist} spherical galactic models were supplemented with a point source potential to effectively account for the central black hole and radiation pressure \citep{Ciotti_2016,Ciotti_2017,Ciotti_2018}.

In this paper we investigate accretion of polytropic matter onto the symmetry axis in the field of a cylindrically symmetric potential. We neglect self-gravity and viscosity and assume the flow velocity is horizontal. This model could be used to approximately describe the accretion of an extended gaseous cloud in the field of an elongated mass concentration, when the  increase in mass of the object could be neglected. Under cylindrical symmetry the interesting issue is the rotation of the flow, and thus it is necessary to consider accretion with non-zero angular momentum. We call this kind of accretion a {\it spiralling accretion} to distinguish it from purely radial accretion onto the symmetry axis which we also investigate here as a particular reference model. 

Physically, the idealized cylindrical model describes accretion of gas onto elongated concentrations of matter such as filaments or arms -- the structures were described in the astrophysical literature by infinite 
cylinders (we  discuss this issue below in more detail). The only difference is
that we do not consider a self-gravitating collapse of infinite massive cylinders, but only accretion of the surrounding medium onto such structures. We recall also that the particular model can be arrived at by considering general accretion disks in spherical potentials -- the cylindrical approximation is fine if the disk is thin at any radius. As shown in the Appendix, the equations obtained  in the leading order of perturbations for general flows reflection-symmetric with respect to the equatorial plane are the same as for the spiralling accretion model.

{\it Astrophysical motivation behind considering cylindrically symmetric purely radial and non-radial horizontal flow models.} The spiralling accretion model and its purely radial version considered in this paper, belong to a class of horizontal flow models under full cylindrical symmetry considered in various contexts of astrophysics. Various such flows or even static equilibria of infinite cylinders have been intensively studied over several decades in view of the interests in description of approximately cylindrical structures of astronomical objects such as filaments, slender rings and arms, in various astrophysical contexts (filaments of gas clouds, arms of spiral galaxies) even on cosmological scales (filamentary forms of clusters of galaxies). The cylindrical models are natural counterparts of spherically symmetric models solved with the same methods.
\cite{1942ApJ....95...88R} used infinite cylinder model to develop the theory of the equilibrium of gaseous rings modelled as toroids (of the same cross-sectional radius), applicable in situations when the cylinder's radius is much lower than the central radius of the toroid.
Two decades later, in the context of gaseous rings, \cite{1964ApJ...140.1056O} formulated -- what we might loosely call -- a cylindrical version of spherical stars. He investigated infinite self-gravitating polytropic cylinders (as the leading approximation of slender rings) for the purpose of  finding the equilibrium distributions of pressure, density, and gravitational potential. He derived the cylindrical version of the Lane-Emden equations known from the theory of spherical stars and obtained a few exact solutions, in particular, for a finite radius homogeneous liquid cylinder or infinite radius isothermal perfect gas cylinder, both with finite mass per length. Next, he extended these investigations to true rings \citep{1964ApJ...140.1067O}. 
This illustrates the direction of developing astrophysical models - one starts with idealized models not realised in nature as a background to understand more realistic models, and so to see in which astrophysical situations the idealized models can, nevertheless, be still used as an approximation.

\cite{1953ApJ...118..116C} pointed out that astrophysical objects such as a spiral arm can be idealized as an infinite cylinder. 
Under the assumption of all quantities depending solely on the distance from the axis of the cylinder, among other things, they derived the virial theorem for an infinite cylindrical distribution of matter with uniform magnetic field, and
investigated radial pulsations of an infinite cylinder along the axis of which the magnetic field is acting, and also stability of the cylinder for transverse oscillations in the incompressible case. Within this framework, they showed strong stabilizing effect of magnetic fields. In particular, taking as typical of a spiral arm of a galaxy, a value of $250$ parsecs for the cylinder radius and  density of $2\times 10^{-24}$ gram per cm$^3$, they
predict magnetic field in excess of $7$ microgauss to effectively remove instabilities. It is striking that the idealized cylindrical model predicts magnetic fields of the same order as met in galaxies (it was recently found that fields of this order can also explain the observed difference between slower stellar rotation relative to gas rotation in the outer part of Galaxy \citep{JalochaBratek2016}).
In the same context of gravitational stability of spiral arms  \cite{1963AcA....13...30S} 
considered an infinite isothermal cylinder, extending investigations to other cylindrically symmetric magnetic fields like, for example, fields with  concentric circular magnetic lines around the symmetry axis, under the same assumption that all quantities are functions of only the radial distance from the symmetry axis.

\indent
Models exploiting full cylindrical symmetry of infinite cylinders are used also to understand fragmentation instabilities within  elongated structures of astrophysical interest. 
Interstellar clouds have very complex structure, often consist of elongated, filamentary structures containing stars and dense molecular clouds, and  fragmented into globule-like beads \citep{1979ApJS...41...87S, 1984A&A...137...17G, 1987ApJ...321..855H}. Initially elongated structures arising, as it seems, by the preceding fragmentation of sheet-like interstellar clouds, become elongated more and more as they go on collapsing and mostly end as very slender cylinders, as strongly suggested by the non-linear stability analysis of an infinite isothermal gas layer performed by \cite{1987PThPh..78.1051M}. In particular,  based on the exact infinite cylinder equilibrium solution obtained by \citep{1964ApJ...140.1056O}, the authors gave simple criterion for the equilibrium and radial collapse of an infinitely long self-gravitating isothermal finite-radius cylinder with constant external boundary pressure  to understand fragmentation instabilities of the filamentary clouds (under the assumption of all the quantities being independent of $z$). They presented also a self-similar solution (involving time and the radial distance from the symmetry axis) which describes the cylindrically-symmetric collapse onto the cylindrical axis of an isothermal cylindrical cloud  assumed to be infinitely long in the direction of the axis, and which, at any moment, is identical
to the \citet{1964ApJ...140.1056O} equilibrium solution. A uniformly axially magnetized version of this problem was considered by 
    \cite{1987PThPh..77..635N} where gravitational instability of the cylinder was investigated in linear approximation by perturbing the same equilibrium solution and showing that the magnetic field has a stabilizing effect.
Self-similar solutions for collapsing self-gravitating infinite isothermal cylinders were investigated further, for example \citep{1992ApJ...388..392I}. In the same similarity approach \citet{1998PASJ...50..577K} consider solutions with arbitrary polytropic index, then \citet{2009PASP..121..485H}
 extended results to dynamic polytropic index. Self-similar collapse of magnetized isothermal cylinders was investigated in 
\cite{2003ApJ...593..426T}. \citet{2016MNRAS.456L.122L} consider more general magnetic fields compatible with the cylindrical symmetry of infinite cylinders (such as helical fields).  

Although the spiralling accretion model provides only approximate description, the usefulness of the model in astrophysical situations seems clear. Moreover, solutions to simple models find practical use in testing numerical codes. For example, on the occasion of developing  magnetohydrodynamical simulation codes in cylindrical geometry, \citet{skinner2010} recover what they call a cylindrical version of Parker's spherically symmetric wind \citep{parker1958} using an adiabatic exponent $5/3$ in a power law-potential inversely proportional to the distance from the symmetry axis. In this case, they used contours of the mass flux as a test for implementing cylindrical coordinates in the simulation codes. 

As the approximating potentials in this paper we consider a class of power-law potentials $\frac{x^{1-\beta}}{1-\beta}$ enumerated with a single parameter $\beta>1$ and complete it by adding a logarithmic potential, which is natural to consider under cylindrical symmetry or as a limit $\beta\to1$ of power-law potentials ($x^r\sim 1+r \ln{x}$ as $r\to0$), here $x$ is the radial distance from the axis of the cylindrical symmetry in units defined later. 
The approximating potential needs not to be interpreted as entirely of the gravitational origin and may effectively include also some non-gravitational phenomena in the accretion process. As an example in the context of spherical Bondi accretion model, on can consider the radiation pressure due to scattering of electrons. It is known that in the optically thin regime, the radiation can be easily accounted for at low accretion rates as a correction that effectively reduces the amplitude of the point mass potential in Bondi model by some factor \citep{Ciotti_2016}. 
Considering the class of simple potentials allows to see in a single setup how the flow picture is affected by the shape of the potential and the value of the polytropic exponent. The accretion problem as formulated is simple. The resulting equations are easily tackled with. They define a dynamical system of which 
solutions can be studied qualitatively and understood in analytical way. Unlike spherical accretion models referred to above, the spiralling accretion model seems to have not been investigated in more detail so far.   

We also investigate the simpler case of purely radial accretion. Cylindrical analogues of spherical accretion or collapse are considered in the literature. A fully cylindrically-symmetric accretion model of infinite self-gravitating cylinder of polytropic material was recently considered in \cite{lou2016} where dynamical solution were obtained by the similarity argument.  This work is the cylindrical analogue of the collapse of a spherically-symmetric polytropic gaseous cloud to which the similarity method was applied \cite{1985ApJ...293..494B}. Although cylindrical accretion without angular momentum may seem too simplistic from the astrophysical standpoint, it is interesting as an example of application of the mathematical tools we use in this paper. The radial accretion model provides also a natural reference background for the purpose of comparison with the spiralling accretion model. We find it easier to investigate the radial accretion model first and then obtain analogous results for the spiralling accretion model.
The purely radial limit of our spiralling accretion model has also some other merits. Both the Bondi radial accretion model, and the cylindrical accretion model with or without angular momentum, belong to a class of effectively one-dimensional models of stationary flow. They define their corresponding Hamiltonian systems which can be studied with the help of the same mathematical tool. As we shall see, the Bondi model and the cylindrical radial accretion model share similar types of phase portraits on the phase plane (albeit with different interpretation of the phase variables) that can be classified by the signature of some Hessian matrix, which may be of the hyperbolic, parabolic or elliptic type. The critical values of the free model parameters that demarcate various types of portraits are different in each model. The Hessian determinant, so to say,  detects some critical signatures of the flow. For example, as we will find, for the classical Bondi model critical is the polytropic exponent value of $\alpha=5/3$ at which the Hessian determinant vanishes and changes sign. We may recall that solutions with $\alpha>5/3$ were not considered by Bondi and they are regarded as unphysical \citep{shu1991}.
We find that the critical polytropic exponent is different in the cylindrical radial accretion model. It is variable, lower or greater than $5/3$, depending on the exponent $\beta$ in the potential term. In particular, the phase portrait with the stationary hyperbolic point observed for the classical Bondi accretion with $\alpha<5/3$ (which includes physical solutions) is possible also with $\alpha>5/3$ in the case of the cylindrical accretion. Similarly, the phase portrait with the stationary elliptic point that comprises solutions we would consider unphysical in both models, is also possible with $\alpha<5/3$ in the case of cylindrical accretion. 

An interesting feature of the cylindrical accretion model with $\alpha=1$ and any $\beta$, is that it belongs to a class of models exactly solvable in terms of the Lambert $W$ function, both for purely radial and spiralling accretion. More generally, it would be so for any other accretion or wind problem reducing in the isothermal case to a unifying integral of the general form $y^{-m}\,\mathcal{K}(x)+\ln{y}-\mathcal{E}(x)=0$, and this is the case for spiralling accretion in any radius-dependent potential (here $y$ represents the density as a function 
of some spatial variable $x$ (e.g. the radial distance), while $\mathcal{K}(x)$ and $\mathcal{E}(x)$ represent, respectively, the contribution from the kinetic 
term determined by the assumed symmetry of the flow, and the contribution from the adopted potential term). 
In particular, in the context of accretion or wind processes, it has been already known that exact solutions can be also found for the isothermal radial flow in the Bondi model with the point mass potential \citep{cranmer2004} as well as with other spherical potentials such as Jaffe or Hernquist potentials \citep{Ciotti_2017}. 

Interestingly, in the case of spiralling accretion in the logarithmic potential, exact solutions can be found for arbitrary polytropic exponent. This is particularly interesting for two reasons, first, because the radial accretion onto a thin string provides a cylindrically symmetric counterpart of the classical Bondi accretion onto a point mass, secondly, because this correspondence can be extended to cylindrical accretion with non-zero angular momentum for matter with arbitrary polytropic index.  We give also two non-trivial solutions for the purely radial and spiralling accretion expressible without the use of the $W$ function.

\section{General remarks on the radial and spiralling accretion}

Throughout the text we will be using the following notational conventions. In this model we have three mechanical units (that of time, length and mass) which we may specify by setting three arbitrary and dimensionally independent combinations of them. We have also two dimensional and two dimensionless free parameters of the model describing the polytropic material and the potential. For $\beta>1$ the potential is given by 
$\frac{v_o^2}{1-\beta}\br{R/R_o}^{1-\beta}$.
If we set some length scale $R_o$ then $v_o$ is a free parameter with the dimension of velocity that specifies the strength of the potential force. Next, we introduce some scale of mass density 
$\rho_o$ yet to be specified, 
and express the polytropic equation of state in the form 
$p(R)=p_o\cdot \br{\rho(R)/\rho_o}^{\alpha}$ with free parameter $p_o$ having the dimension of pressure and defining the value of the speed of sound $\sqrt{\alpha\,p_o/\rho_o}$ at density $\rho_o$. The pressure parameter can be alternatively expressed in terms of a dimensionless sonic 
parameter $\upsilon=\sqrt{\frac{p_o}{\rho_o v_o^2}}$. We may now define $\rho_o$ 
so that $\upsilon=1$, provided that $\alpha>1$. When $\alpha=1$ we have additional scaling freedom $p_o\to k p_o$ and $\rho_o\to k \rho_o$ with arbitrary $k>0$ and we cannot fix $\rho_o$ in this way, and so we have to keep the free parameter $\upsilon$ in the equations. Having said this, we may now introduce the following dimensionless quantities
$$x=\frac{R}{R_o},\quad{y(x)}=\frac{\ro(R)}{\ro_o},\quad{}w(x)=
\frac{-v_{R}(R)}{v_o},\quad u(x)=
\frac{v_{\phi}(R)}{v_o},\quad{}\kappa=
\frac{A}{R_o\,\ro_o v_o}, \quad \lambda=\frac{J}{R_o\,v_o}.$$
Here, $v_R$ and $v_{\phi}$ are the radial and the azimuthal components of the flow velocity in cylindrical coordinates,
$\ro$ is the mass density, $A$ is the accretion rate and $J$ is the specific angular momentum. We introduce also the dimensionless speed of sound $c(x)$ expressed in the unit of velocity $v_o$:
$$\quad c^2(x)\equiv
\frac{1}{v_o^2}\frac{p'(R)}{\rho'(R)}=\alpha\,\upsilon^2\,y^{\alpha-1}(x).$$ 
%\medskip
The hydrodynamical equations for a polytropic matter with horizontal flow velocity under cylindrical symmetry (see  equations \eqref{eq:A1} in Appendix), imply in the field of the assumed class of potentials the following first integrals 
\begin{equation}\label{eq:econstraint}
\eps=
\frac{1}{2}w^2+\frac{1}{2}u^2+U(x,y)
,\quad \lambda= x\,u,\quad
\kappa=x\,y\,w,
\end{equation}
with constant parameters: $\eps$ (specific energy), $\lambda$ (specific angular momentum) and $\kappa$ (the accretion rate). The potential $U(x,y)$ is defined as  
\begin{equation}\label{eq:Upotential}
U(x,y)=\left\{\begin{array}{cllc}
\frac{\alpha}{\alpha-1}\,y^{\alpha-1}-\frac{1}{\beta-1}
\frac{1}{x^{\beta-1}},&
\alpha>1, & \beta>1, & 1-\alpha<\gamma/2<\beta-1\\[\smallskipamount] 
\frac{\alpha}{\alpha-1}\,y^{\alpha-1}+\ln{x},&
\alpha>1, & \beta=1, & \gamma<0\\[\smallskipamount] 
\upsilon^2\ln{y}+\ln{x},&
\alpha=1, & \beta=1, & \gamma=0\\[\smallskipamount] 
\upsilon^2\ln{y}-\frac{1}{\beta-1}
\frac{1}{x^{\beta-1}},&
\alpha=1, & \beta>1, & \gamma>0\\[\smallskipamount] 
\end{array}\right.,\qquad \gamma:=4+(\alpha+1)(\beta-3).
\end{equation}
Here, we have introduced a criticality parameter $\gamma$ which we will frequently be using throughout the text. As we will see, this precise combination of exponents $\alpha$ and $\beta$ distinguishes between qualitatively distinct regimes of the accretion model (with the critical value being $\gamma=0$).
We remind that by referring to {\it spiralling} accretion, we mean horizontal velocity field with the azimuthal component satisfying precisely the inverse distance law $u(x)=\lambda/x$.  The vorticity vector of the spiralling flow vanishes, which means that the accretion is locally irrotational.

\begin{figure}[h]
\begin{center}
\centering
\framebox{\includegraphics[width=0.33\textwidth]{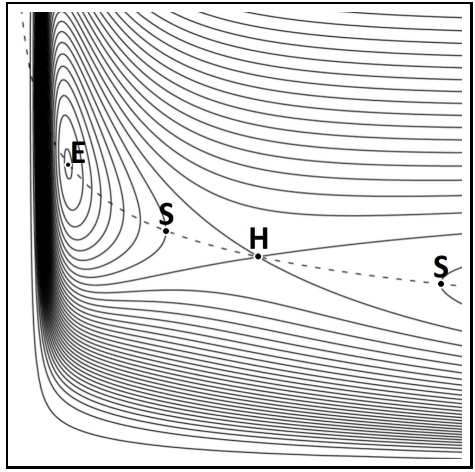}}
\caption{\label{fig:00} 
An example phase portrait for polytropic spiralling accretion on the position-density $(x,y)$ plane (in this work we show also phase portraits on the position-velocity $(x,w)$ plane, which are deformed versions of the position-density phase diagrams obtained by suitable change of coordinates). In this figure, the solid lines are level lines of the Hamiltonian $F(x,y)$. The dashed line is the locus of all shock points (represented by {\sf S} in the diagram); the line represents a solution $y=\mathcal{Y}(x)$ of equation $\partial_yF(x,y)=0$ for all $x$ such that $F_x(x,\mathcal{Y}(x))\neq0$. A level line may intersect with the shock curve at some shock point such as {\sf S} to the left of point {\sf H} (which also can be called a concave turning point) or such as {\sf S}  to the right of {\sf H} (which is then a convex turning point). In this model the pressure gradient exerted on polytropic matter becomes infinite at turning shock points. 
A level line may cross the shock curve also at some isolated points, where besides the constraint $\partial_yF(x,y)=0$, also $\partial_xF(x,y)=0$ holds. We call such points stationary points or critical points of the Hamiltonian. There are two such points in this figure marked with {\sf E} and {\sf H} (there can be $1$, $2$ or $3$ such points in this model). Point {\sf H} is a stationary hyperbolic point -- the level lines are  hyperbolas to second order in variations $\delta{x}$ and $\delta{y}$ from the position $(x,y)$ of {\sf H}, except for two solid lines that cross each other at {\sf H} and correspond to the asymptotes of those local hyperbolas. The two asymptotes locally overlap with two level lines that cross each other at {\sf H} (called separatrices). The two level lines are regular solutions with definite derivatives at {\sf H} -- for this reason we also call {\sf H} a regular sonic point (all shock points {\sf S} are also sonic points in this model).  Point {\sf E} is a stationary elliptic point (the level lines form ellipses to second order in  variations $\delta{x}$ and $\delta{y}$ from the position $(x,y)$ of {\sf E}). The limiting ellipses about {\sf E} degenerate to a point at {\sf E}. 
}\end{center}
\end{figure}

The three integrals of motion in equation \eqref{eq:econstraint} provide three constraints to be satisfied by four variables $u$, $w$, $y$ and $x$. The solutions cannot be given in analytical form 
unless $\alpha$ and $\beta$ attain particular values. Nevertheless, the solutions can be still investigated qualitatively with basic methods. Using the last two integrals, we eliminate the velocities from the first one and consider the following unifying integral
\begin{equation}\label{eq:intsurf1}
F(x,y)\equiv\frac{1}{2}\frac{\kappa^2}{x^2y^2}+\frac{1}{2}\frac{\lambda^2}{x^2}+U(x,y)=\eps,\qquad{}x>0,\quad{}
y>0.\end{equation} 
Regarded as a function of two independent variables, $F(x,y)$ describes an {\it energy surface} over the phase plane $(x,y)$. 
Solutions given in the implicit form $F(x,y)=\eps$ can be then visualized as
level lines on that plane, as sketched in Figure \ref{fig:00} where some example family of solutions with fixed $\kappa$ and $\lambda$ is shown for the purpose of illustration of various phase portrait features referred to in the text.
The problem of finding level lines can be reinterpreted as a Hamiltonian system
with $x$ being the position variable, $y$ the momentum canonically conjugate to $x$, and
$F(x,y)$ playing the role of a Hamiltonian:
$$\dot{x}=\{x,F\}=\partial_yF(x,y),\quad \dot{y}=
\{y,F\}=-\partial_xF(x,y).$$ 
By choosing other pairs of independent variables, other Hamiltonians
would be possible to start with, and later we give examples where it would be more appropriate to consider $x$ and $y$ as functions of $w$ or $x$ and $w$ as functions of $y$, in which case the exact solutions are easier to find.
This method might have turned out fruitful when applied also to other accretion problems.

Since $\partial_t F=0$ for the Hamiltonian \eqref{eq:intsurf1},  $F$ is
conserved for solutions, consistently with the method of level lines. The equation
$F(x,y)=\eps$ can be solved locally for $y$ if
$\partial_yF(x,y)\ne0$. Then $\dot{x}(t)\ne0$, and the integral curve $(x(t),y(t))$ can be 
re-parametrized and represented as $y(x)$, in which case
$y'(x)\equiv-\partial_xF/\partial_yF$ along that curve. 
Breaking the regularity condition $\partial_yF(x,y)\ne0$ at some point  means the occurrence of a {\it density shock}. Wherever $\partial_yF(x,y)=0$ and $\partial_xF(x,y)\ne0$ at some point, then $y'(x)$ diverges at that point, which we therefore call a {\it shock point}. In the neighbourhood of that point, we may represent the solution in a reversed form $x(y)$ parametrized with $y$. 
The locus of all possible shock points lying entirely on the energy surface we call a \emph{shock curve}. Parametrized with $x$  the shock curve reads:
$$y=\mathcal{Y}(x),\qquad z=\mathcal{Z}(x)\equiv F(x,\mathcal{Y}(x)), \quad
\mathrm{provided\ that}\quad 
\partial_xF(x,\mathcal{Y}(x))\neq0,$$
 where 
$$
%p_{sonic}(s):\qquad
%x_s(s)=s,\quad 
 \mathcal{Y}(x)=
\br{\frac{\kappa^2}{\alpha\,x^2}}
^{\frac{1}{\alpha+1}}\quad
%,\qquad s\in(0,\infty)  
\mathrm{for}\ \alpha>1, \quad \mathrm{and} \quad \mathcal{Y}(x)=\frac{\kappa}{\upsilon x} \quad \mathrm{for} \ 
\alpha=1.
$$
Since $\partial_yU(x,y)=y^{-1}c^2(y)$ both for $\alpha>1$ and $\alpha=1$, the radial velocity on the shock curve equals the speed of sound: $c^2(\mathcal{Y}(x))=w^2(x,\mathcal{Y}(x))$. Therefore, we define a {\it sonic curve} as the locus of all points of the phase plane where the accretion velocity equals the speed of sound. Unlike for the shock curve, we also allow points for which $\partial_xF(x,\mathcal{Y}(x))=0$ (outside these isolated points the sonic curve and the image of shock curve on the phase plane overlap with each other).  
The sonic curve divides the phase plane $(x,w)$ into two parts. Above that curve solutions are {\it supersonic}, below that curve solutions are {\it subsonic}, while on the $(x,y)$ phase plane the converse applies. 
%%%%
An example phase portrait on the $(x,y)$ phase plane is shown in Figure \ref{fig:00} together with the locus of all (sonic) shock points (shown with the dashed line) and two representative shock points (marked with {\sf S}). 
%%%%
%%%
%%%%
Note that the total velocity is supersonic at the shock points in the case of spiralling flow. 
The condition $\partial_xF(x,y)\neq0$ in the definition of shock curve is important.
Sonic points where both $\partial_yF=0$ and $\partial_xF=0$ are stationary points of the energy surface (in Figure \ref{fig:00} there are two such points: {\sf E} -- the elliptic stationary point, and {\sf H} -- the hyperbolic stationary point). Although  $\dot{y}$ and $\dot{x}$ both vanish at a stationary point, their ratio  $y'(x)$ 
becomes a  $\frac{0}{0}$  indeterminate expression that may evaluate to a finite number regarded as a limit. In the case of a hyperbolic stationary point, the solution $y(x)$ crossing the sonic curve at that point has finite derivative $y'(x)$ and the stationary point should be considered as a {\it regular sonic point} (in Figure \ref{fig:00} there are two level lines intersecting at the hyperbolic stationary point {\sf H} representing two independent smooth solutions passing through that point).

\section{Qualitative analysis of solutions ($\alpha>1, \ \beta>1$)\label{sec:qualitative}}

The form of the Hamiltonian allows to express solutions in terms of known functions only in particular cases. Therefore, we focus more on the qualitative properties of solutions for $\alpha>1$ and $\beta>1$. 
There are several 
useful identities satisfied by $\mathcal{Z}(x)$. The first is the inhomogeneous second order differential equation
$$\left( \alpha +1 \right) \,x^2\,\mathcal{Z}''(x) + 
   \left( 5\,\alpha  - 3 + \gamma  \right) \,x\,\mathcal{Z}'(x) + 
   2\,\left( \alpha  - 1 \right) \,\left( \beta  - 1 \right) \,
    \mathcal{Z}(x) =2\,\left( 3 - \beta  \right) \,
    \frac{{\lambda }^2}{x^2}.$$ We arrive at it just by considering
a linear combination of a general function $\mathcal{Z}$ and its derivatives $\mathcal{Z}'$ and $\mathcal{Z}''$, form-invariant with  respect to the scaling transformation $x\to k\, x$, that is, $a\,\mathcal{Z}(x)+ b\, x\,\mathcal{Z}'(x)+c\, x^2\mathcal{Z}''(x)$,  and then substituting the particular form $\mathcal{Z}(x)=F(x,\mathcal{Y}(x))$ and choosing coefficients $a,b,c$ so that only a term involving $\lambda$ survives. Now we may forget about the particular form of $\mathcal{Z}$.
The method of solving this differential equation distinguishes the case when there is a multiple solution to the discriminant equation for the characteristic roots or when such a root equals the exponent in the inhomogeneous term (the characteristic roots are $r=1-\beta<0$ and $-2< r=-2\frac{\alpha-1}{\alpha+1}<0$). The first case occurs for $\alpha$ and $\beta$ constrained by the condition $\gamma=0$. Note, that the particular $\mathcal{Z}(x)=F(x,\mathcal{Y}(x))$ does not involve the independent solution $\frac{\ln{x}}{x^{\beta-1}}$ possible in this case. The second case occurs when 
$\beta=3$ (or $\gamma=4$). Then the inhomogeneous term disappears from that equation, however the centrifugal term is still present in the general integral $\mathcal{Z}$.  There is also a third case $(\alpha-1)(\beta-1)=0$ when one of the roots or both are zero. The three possibilities will be investigated separately. 
Next two identities are satisfied by the invariants of the Hessian matrix evaluated on the shock curve:
$$\det{[\partial^2_{ij}F(x,y)]}\left|_{y=\mathcal{Y}(x)}\right.=\frac{\alpha  + 1}{{\kappa }^2\,x^2}\,
  {\left( \alpha \,{\kappa }^{\alpha  - 1}\,x^2 \right) }^{\frac{4}{\alpha+1 }}\,
  \mathcal{Z}''(x)$$
  $$
  \tr{[\partial^2_{ij}F(x,y)]}\left|_{y=\mathcal{Y}(x)}\right.=\mathcal{Z}''(x) + \frac{2}{\kappa }\,{\left( \frac{\alpha \,{\kappa }^{\alpha  - 1}}
       {x^{\alpha  - 1}} \right) }^{\frac{3}{\alpha+1  }}\,
   \left( \Xi + \frac{1}{\Xi} \right),\qquad
         \Xi=\frac{\alpha+1 }{2}\,{\left( \frac{\alpha }{{\kappa }^2}\,
          x^{\alpha +3} \right) }^{\frac{1}{\alpha  + 1}},
  $$
where $\partial^2_{ij}F(x,y)$ is shorthand notation for the Hessian matrix
  $$\partial^2_{ij}F(x,y)\equiv \left[\begin{array}{ll}\partial^2_{xx}F&\partial^2_{xy}F\\
\partial^2_{yx}F&\partial^2_{yy}F\end{array}
\right].$$  
  It follows from the above identities that: a local maximum of $\mathcal{Z}$ is a hyperbolic stationary point of the Hamiltonian, while a local minimum of $\mathcal{Z}$ is an elliptic stationary point of the Hamiltonian. One can also infer that the only non-zero eigenvalue of the Hessian is positive at the inflection point of $\mathcal{Z}(x)$ at which $\mathcal{Z}''(x)=0$. If  additionally $\mathcal{Z}'(x)=0$ at that point (stationary inflection point), the corresponding stationary point of the Hamiltonian is parabolic. For later use, we give also the following identities 
\begin{equation} \label{eq:id2}
         \mathcal{Z}'(x) + \frac{2}{x}\,\mathcal{Z}(x) = 
  \frac{\beta  - 3}{\beta  - 1}\,
    \frac{1}{x^{\beta }} + 
\frac{2}{\alpha  - 1}\,
      {\left( \alpha{\kappa}^{\alpha  - 1} \right)
          }^{\frac{2}{\alpha+1 }}   x^
      {-\frac{3\,\alpha  - 1}{\alpha+1 }},\qquad
      %\label{eq:id1},
\mathcal{Z}''(x) + \frac{3}{x}\,\mathcal{Z}'(x) = 
  \frac{3 - \beta }{x^{\beta  + 1}} - 
   \frac{4}{\alpha+1 }\,
      {\left( \alpha \,{\kappa }^{\alpha  - 1} \right)
          }^{\frac{2}{\alpha+1 }}
   x^
      {-\frac{4\,\alpha }{\alpha+1 }}.
     \end{equation}

\medskip
\noindent
{\it Turning points}. The shock curve is also the locus of turning points (which we  regard as terminating points of the solutions $y(x)$).  Let us investigate the local behaviour of the inverse solutions $x(y)$ in the vicinity of a shock point.
It suffices to consider only the second derivative $x''(y)$, because $\partial_yF=0$ and $\partial_xF\ne0$, and so
$x'(y)=0$ at the shock point. In this case we obtain $x''(y)|_{\mathcal{Y}(x)}=-\br{\partial^2_{yy}F/\partial_xF}|_{\mathcal{Y}(x)}+(\dots)$ where we have neglected terms vanishing on the shock curve denoted with dots. From this we infer that the curvature of an integral curve crossing the shock curve may change its sign
      only at a stationary point since $\partial^2_{yy}F>0$ on the shock curve. For non-stationary points, we can determine the sign of $x''(y)$ at a given turning point from 
\begin{equation}\label{eq:curvgen}
x''(y)\left|_{y=\mathcal{Y}(x)}
\right.=-\frac{\partial^2_{yy}F(x,y)}{\partial_x F(x,y)}\left|_{y=\mathcal{Y}(x)}
\right.={\left( \alpha \,{\kappa }^{\alpha  - 1} \right) }^
   {\frac{4}{\alpha+1 }}\,
  \frac{\alpha +1}{{\kappa }^2}\,
  \frac{x^{\frac{\alpha  + \gamma+5 }{\alpha +1}}}
   {x^{\frac{\gamma }{\alpha+1 }}\,
      {\left( \alpha \,{\kappa }^{\alpha  - 1} \right)
          }^{\frac{2}{\alpha+1 }} + 
     x^{\frac{\gamma  - 4}{\alpha+1 }}\,
      {\lambda }^2 - 1}.
      \end{equation}
      The shock point is either convex or concave turning point if this sign is positive or negative, respectively. In other words, if $x(y)$ is convex at the shock point, then the subsonic and the supersonic branch of $x(y)$ converge one toward another with decreasing radius in some right neighbourhood of that point. If $x(y)$ is concave then the two branches diverge one from another with decreasing radius in some left neighbourhood of the shock point. If there is no an isolated stationary point, all of the shock points are either left or right turning points.       
            
\smallskip\noindent
{\it The asymptotics}.
We may establish the following inequality satisfied by solutions
\begin{equation}\label{eq:inequ}\frac{1}{2}\,\frac{\alpha  + 1}{\alpha  - 1}\,
   {\left( \alpha \,{\kappa }^{\alpha  - 1} \right) }^
    {\frac{2}{\alpha+1 }}\leqslant
  \frac{1}{
      \beta  - 1 }\,x^{-\frac{\gamma }{\alpha+1 }} + 
   x^{\frac{2\,\left( \alpha-1  \right) }
       {\alpha+1 }}\,\epsilon  - 
   \frac{{\lambda }^2}{2}
  x^{-\frac{4}{\alpha+1 }}.
  \end{equation} We prove it by noticing that for a fixed $x$ the $y$-dependent part in $F(x,y)$ is bounded from below by its value on the shock curve. If the inequality is violated at some $x$, the level line does not cross the line of constant $x$ on the phase plane. 
In this model the solutions may be spatially bounded or extend out to infinity. In the latter case, of interest is the behaviour of the solutions in the limit $x\to+\infty$, and we may distinguish three cases
depending on the sign of $\eps$:
\begin{enumerate}
\item For $\epsilon<0$ no solution may extend out to infinity, because the term on the right of the above inequality diverges to $-\infty$. 
\item For $\eps=0$ solutions may extend out to infinity provided either $\gamma<0$, or both  $\gamma=0$ and $\alpha\kappa^{\alpha-1}\leqslant 1$.
 Since the positive terms in $F(x,y)$ must be all zero if $\eps=0$, then both $y\to0$ and $w\to0$, such that $x\,y\to\infty$ as $x\to\infty$. 
 \item 
For $\eps>0$ the asymptotics of the solutions extending out to the infinity can be found in the following way. We suppose that $w^2=0$ or  $w^2>0$ at infinity and then we infer  that 
$y=0$ or $y=y_{\infty}>0$. In the former case we assume $y\sim x^{-r}$ and in the latter that $y\sim y_{\infty}(1+u)$, and expand to second order, assuming $|u|\ll1$. This way we find that  for $\eps>0$ 
the density function behaves at infinity in two ways. There is a supersonic branch for which
\begin{equation}\label{eq:supasympt}
y(x)\sim \frac{|\kappa|}{x\sqrt{2\eps}}, \qquad w(x)\sim\sqrt{2\eps}\end{equation}
 and a subsonic branch for which 
 
 \begin{equation}\label{eq:subasympt}
\large 
y(x)\sim{\left( \frac{\alpha  - 1}{\alpha }\,\epsilon  \right)
      }^{\frac{1}{\alpha  - 1}}
      \left\{\begin{array}{ll}
 1 + \frac{1}
      {\left( \alpha  - 1 \right) \,
        \left( \beta  - 1 \right) \,\epsilon }\,
     \frac{1}{x^{\beta  - 1}}, 
     &1<\beta<3     \\[\medskipamount] 
     1 + \frac{1 - {\lambda }^2 - 
      \frac{{\kappa }^2}
       {{\left( \frac{\alpha  - 1}{\alpha }\,
            \epsilon  \right) }^{\frac{2}{\alpha  - 1}}
         }}{2\,\left( \alpha  - 1 \right) \,\epsilon }
         \,\frac{1}{x^2}, & \beta=3 \\[\medskipamount] 
    1 - \frac{{\lambda }^2 + 
      \frac{{\kappa }^2}
       {{\left( \frac{\alpha  - 1}{\alpha }\,
            \epsilon  \right) }^{\frac{2}{\alpha  - 1}}
         }}{2\,\left( \alpha  - 1 \right) \,\epsilon }
         \,\frac{1}{x^2} & \beta>3  \end{array}\right., \qquad w(x)\sim\frac{\kappa}{x}\br{\frac{\alpha}{(\alpha-1)\eps}}^{\frac{1}{\alpha-1}}.
\normalsize         
         \end{equation}
 \end{enumerate}

\noindent
{\it Solutions in the proximity of the centre}.
With the use of the inequality \eqref{eq:inequ} we may draw conclusions also in this region. For $\gamma<0$ neither purely radial nor spiralling solutions are possible below some finite radius. The radial flow may reach the center (that is, the axis of the
cylindrical distribution $x$=0) if $\gamma>0$ or when $\gamma=0$ at small enough accretion rate $\alpha \kappa^{\alpha-1}\leqslant 1$. 
As to the spiralling flow there are three cases which we consider: a) for $0\leqslant\gamma<4$ and $x$ small enough the centrifugal term in \eqref{eq:inequ} is dominating and repels matter from the centre; b) for $\gamma=4$ matter may reach the centre only for angular momenta low enough, such that $\lambda^2<1$; c) for $\gamma>4$ the centrifugal term is too weak compared with the attractive term for any $\lambda$ and matter may reach the centre.

One can make also some predictions concerning the way the solutions approach the center. As an example we consider the subsonic branch of the radial accretion. In this case the limiting solution can be also predicted  perturbatively. For this purpose it is easier to consider the Hamiltonian $F(x,\frac{\kappa}{x\,w})$ expressed in terms of phase variables $(x,w)$.
We substitute the ansatz $w(x)\sim a x^{s}(1+b x^r)$  and find the 
unknown parameters in the linear approximation with respect to parameter $b$, assuming $s>0$ an $r>0$.
 The resulting limiting expressions for the subsonic branch reads 
$$y(x) \sim \frac{{\left( \frac{1}{\alpha }\,
         \frac{\alpha  - 1}{\beta  - 1} \right) }^{\frac{1}{\alpha  - 1}}}
      {x^{\frac{\beta  - 1}{\alpha  - 1}}}\,
   \left( 1 + \epsilon \,\frac{\beta  - 1}{\alpha  - 1}\,
      x^{\beta  - 1} \right),\qquad
      w(x) \sim \kappa \,\frac{x^{\frac{\beta  - \alpha }{\alpha  - 1}}}
    {{\left( \frac{1}{\alpha }\,\frac{\alpha  - 1}{\beta  - 1} \right) }^
      {\frac{1}{\alpha  - 1}}}\,
   \left( 1 - \epsilon \,\frac{\beta  - 1}{\alpha  - 1}\,
      x^{\beta  - 1} \right), \qquad
\lambda=0, \quad \gamma>0 \quad (x\searrow0).$$
We have verified that the second order term in the expansion method tends to $0$ if $\beta>\alpha$, and no reservation was necessary for the upper bound of $\gamma$. 
It follows from the limiting solutions, that the density $y(x)$ always diverges as $x\to0$,  while the radial velocity $w(x)$ converges to $0$ for $\beta>\alpha$ or to a finite value $\kappa\,{\delta}^{\frac{1}{\delta-1}}$ for $\alpha=\beta=\delta>1$, and diverges to $+\infty$ for $\alpha>\beta$.

\section{Radial accretion in the power-law potential \  ($\alpha\geqslant1,\ \beta>1$)}

In what follows we present a more detailed analysis of solutions in the simpler case of radial accretion. Then we will describe similar results for spiralling accretion to see what change to the phase diagram pictures of the radial flow is introduced by the centrifugal term.

\subsection{\label{sec1} Polytropic exponent $\alpha>1$}

\noindent
With the star sign we denote the quantities at the stationary point of the Hamiltonian:
$$\xc=\frac{1}{{\left( \alpha \,{\kappa }^{\alpha  - 1} \right) }^{\frac{2}{\gamma }  }},\quad \yc:=\mathcal{Y}(\xc)=\kappa \,{\left( \alpha \,{\kappa }^{\alpha  - 1} \right) }^{\frac{3 - \beta }{\gamma }},\quad \zc:=\mathcal{Z}(\xc)=\frac{\gamma}{2}\,\frac{\br{\alpha\kappa^{\alpha-1}}^{\frac{2(\beta-1)}{\gamma}}}{(\alpha-1)(\beta-1)},\qquad \gamma\ne0.$$
For $\alpha>1$, $\beta>1$ and $\lambda=0$ the shock curve is defined for 
$x\neq\sc$ and reads: 
\newcommand{\rplcd}[2]{#2}
\begin{equation*}
\left.\begin{array}{l} 
	\!\! \rplcd{X(x)=x,\quad
	Y(x)}{y=\mathcal{Y}(x)},
	%	\qquad 
	%\\
	\qquad \mathcal{Z}(x)=
	\left\{\begin{array}{l}
\lrg{\frac{1}{2}\frac{\alpha+1}{\alpha-1}\br{\alpha\kappa^{\alpha-1}}^{
\frac{2}{\alpha+1}}x^{-2\frac{\alpha-1}{\alpha+1}}-\frac{1}{\beta-1}
x^{-(\beta-1)}}\quad\mathrm{for}\ \gamma\ne0,\\[\medskipamount] 
\lrg{\frac{1}{2}\frac{\alpha+1}{\alpha-1}\sq{\br{\alpha\,
\kappa^{\alpha-1}}^{\frac{2}{\alpha+1}}-1}
x^{-2\frac{\alpha-1}{\alpha+1}}}\quad
\mathrm{for}\ \gamma=0\,\wedge\,\alpha\,
\kappa^{\alpha-1}\neq1.
 			\end{array}\right.
	\end{array}\right.
\end{equation*}

\noindent
For $\alpha$ and $\beta$ being fixed, the energy $\zc$ is a function of the accretion rate $\kappa$. Similarly, regarding $\eps$ as a parameter, we may define a {\it critical accretion rate} $\kappac$ obtained by inverting the previous formula for $\zc$:
\begin{equation}\label{eq:crit_accretion}\kappac=\frac{1}{\sqrt[\alpha-1]{\alpha}}\br{\frac{\eps}{\Gamma}}^{\Gamma},\quad \Gamma=\frac{\gamma}{2(\alpha-1)(\beta-1)},\qquad \br{\sgn{\eps}=\sgn{\gamma}}.\end{equation} 

\noindent
As we have noticed earlier, shock points coincide on the phase plane with  sonic points. This happens for $\gamma\ne0$ and for $\gamma=0\,\wedge\,
\alpha\kappa^{\alpha-1}\neq1$. 
For $\gamma=0$ and $\alpha\kappa^{\alpha-1}=1$ there is a simple solution which overlaps with the sonic curve:
\begin{equation}\label{eq:regsol}y(x)={\alpha^{-\frac{1}{\alpha-1}}x^{-\frac{2}{\alpha+1}}},\quad \kappa={\alpha^{-\frac{1}{\alpha-1}}},\quad \beta=\frac{3\,\alpha-1}{\alpha+1}, \quad \eps=0.\end{equation}
This is a regular sonic solution describing radial sonic accretion with no shocks (despite the fact that
$\partial_yF=0$ for that solution).
Coming back to general exponents $\alpha$ and $\beta$, the expression \eqref{eq:curvgen} for $x''(y)$ on the shock curve simplifies to $$\left.x''(y)\right|_{y=\mathcal{Y}(x)}=\frac{\alpha+1}{\kappa^2}
\br{\alpha\kappa^{\alpha-1}}^{\frac{4}{\alpha+1}}
\left\{\begin{array}{cl}
\lrg{\frac{x^{\frac{\gamma+\alpha+5}{\alpha+1}}}{\br{\frac{x}{\sc}}^{\frac{\gamma}{\alpha+1}}-1}},
&\gamma\ne0\\[\bigskipamount] 
\lrg{\frac{x^{\frac{\alpha+5}{\alpha+1}}}{\br{\alpha\kappa^{\alpha-1}}^{\frac{2}{\alpha+1}}-1}},&\gamma=0\,\wedge\,
\alpha\kappa^{\alpha-1}\neq1.\end{array}\right.$$
Accordingly, 
 for $\gamma<0$, the function
$x(y)$ representing a solution in the neighbourhood of the shock point where $x(y)$ intersects the shock curve, is convex for $x<\sc$ or concave for $x>\sc$,  and for
$\gamma>0$ the converse is true.

The intersection points of the shock curve with the plane of constant specific energy, determine the position of shock sonic points of a given level curve (corresponding to positive roots of the equation $\mathcal{Z}(x)=\eps$). The number of such points depending on $\eps$
can be easily established based on the behaviour of function $\mathcal{Z}(x)$. For $\gamma\ne0$, the $\xc$ is the only local extremum of $\mathcal{Z}(x)$:
$$\mathcal{Z}(\xc)=\zc,\;\; \mathcal{Z}'(\xc)=0,\;\; \mathcal{Z}''(\xc)=-\gamma\,\frac{\br{\alpha\, \kappa^{\alpha-1}}^{\srg{\frac{2(1+\beta)}{\gamma}}}}{\alpha+1},\quad \gamma\ne0. $$
The point is a
local minimum ($\zc<0$ for $\gamma<0$) or a local maximum ($\zc>0$ for $\gamma>0$). The extrema of $\mathcal{Z}(x)$ are global, because 
for $\gamma<0$: $\mathcal{Z}(x)\nearrow+\infty$ as $x\searrow0$ and $\mathcal{Z}(x)\nearrow0$ as $x\nearrow+\infty$, while for $\gamma>0$: $\mathcal{Z}(x)\searrow-\infty$ as 
$x\searrow0$ and $\mathcal{Z}(x)\searrow0$ as $x\nearrow+\infty$. For $\gamma=0$ and 
$\alpha\kappa^{\alpha-1}\ne1$ there are no extrema and $\mathcal{Z}(x)$ is monotone with constant sign: 
for $\alpha\kappa^{\alpha-1}<1$ the function diverges to $-\infty$ as $x\searrow0$, and $\mathcal{Z}(x)\nearrow0$ as $x\nearrow+\infty$;
while for  $\alpha\kappa^{\alpha-1}>1$ the function diverges to $+\infty$ as $x\searrow0$ and $\mathcal{Z}(x)\searrow0$ as $x\nearrow+\infty$. 
 From this analysis it follows that except for the regular sonic solution shown in equation \eqref{eq:regsol}, the number of shock sonic points can be $0$, $1$ or $2$, depending on the shape of the energy surface and the position of the sectional plane of constant specific energy.

So far we have considered solutions close to shock
points. By examining the geometry of the
energy surface we can make an insight into the global
structure of solutions. Crucial is the shape of the surface in the neighbourhood of the stationary point. The shape can be inferred based on the eigenvalues of the Hessian matrix evaluated at that point. Accordingly, with the help of the following result\begin{eqnarray}
&&\partial_xF=0,\quad
\partial_yF=0,\quad
\partial^2_{xx}F=
\frac{4-\gamma}{\alpha+1}\br{\alpha\kappa^{\alpha-1}}^{
\frac{2(1+\beta)}{\gamma}},\nonumber\\ &&
\det{[\partial^2_{ij}F(x,y)]}=-\frac{\gamma}{\kappa^2}
\br{\alpha\kappa^{\alpha-1}}^{\frac{6(\beta-1)}{\gamma}}\qquad
\mathrm{at}\quad x=\xc,\,y=\yc, \label{eq:hess1}
\end{eqnarray}
we can distinguish three cases (remembering that $\alpha>1$ and $\beta>1$):
\newline
\noindent
{\it i).} For $\gamma<0$ the stationary point is an elliptic point, in this case it is a local minimum below zero, $\zc<0$. Then, for $-|\zc|<\epsilon<-|\zc|+\delta<0$ and
$\delta>0$ small enough, the level lines are loops (shrinking to a
point as $\delta\to 0$) with turning points at shock sonic points.  For $\eps>0$ there is only one shock sonic point which is a convex turning point with two solutions extending out to spatial infinity -- one subsonic the other supersonic.
\newline
\noindent
{\it ii).}
For $\gamma>0$ the stationary point is a hyperbolic (or saddle) point with positive energy $\zc>0$. 
Because $\xc$ is the global maximum on
the shock curve for $\gamma>0$, the energy surface
is curving downward in both directions along that curve (and upward in the perpendicular direction).
Therefore, for energies lower than the maximum ($0<\zc-\delta<\eps<\zc$) the
integral curves with turning points --
concave on the left and convex on the right of $\xc$ -- form two disjoint branches of regular solutions crossing the shock curve on the opposite sides of $x=\sc$ and terminating at
their respective  shock sonic points. For energies higher than the maximum ($\zc<\eps<\zc+\delta$) there are no turning points. Then there is 
a subsonic regular solution and a regular supersonic solution, both extending from $x=0$ out to $x=\infty$. 
When $\eps=\zc$, we obtain
two separatrices as regular solutions which cross each other
at the stationary point (which is a regular sonic point) and changing their character at that point from regular subsonic to regular supersonic accretion (or vice versa). 
For $\epsilon<0$ there is only one shock point which is a concave turning point with two solutions extending toward the center on the left of that point, one subsonic and the other supersonic. The position $x_o$ of the single shock sonic point on the sectional plane in the limit as $\eps\nearrow0$,  is given by
$$x_o=\frac{\sc}{\sq{1+\frac{\gamma}{2(\alpha-1)}}^{\srg{{\frac{\alpha+1}{\gamma}}}}}<\sc,\quad \beta>1.$$
 As an example, we can consider an analytical solution with shock points and with features of the phase diagram characteristic for the $\gamma>0$ case. For $\alpha=\delta=\beta$, $\delta>1$ ($\gamma=(\delta-1)^2>0$), the solution can be found in a parametric form, with the radial velocity $w$ playing the role of the free parameter enumerating points along the integral curves: 
   \begin{equation}\label{eq:exactsoldelta} x(w) = \frac{\kappa}{w}\,{\left( \frac{\delta \,
         - \br{{w}/{\kappa}}^{\delta-1}}{
        \left( \delta  - 1 \right) \,
        \left( \epsilon  - \frac{1}{2}\,w^2 \right) } \right) }^
   {\frac{1}{\delta  - 1}},\qquad y(w)=\frac{\kappa}{w\,x(w)}.\end{equation}     
  Since $\gamma>0$ for this solution, the corresponding phase diagram possesses a stationary hyperbolic point corresponding to energy $\zc=\frac{1}{2}\kappa^2\,\delta^{2/(\delta-1)}$ with associated two regular global solutions crossing with the shock curve at that point (the solutions correspond to the $0/0$ singularity of the above expression for $x(w)$, and one of them attains the simple form $y(x)=\br{x\sqrt[\delta-1]{\delta}}^{-1}$).

\noindent
{\it iii).}
For $\gamma=0$ there is only a single shock sonic point present if $\alpha\kappa^{\alpha-1}\ne1$, located at $x=x^{\star\star}$, where
$$x^{\star\star}=\sq{\frac{\br{\alpha\kappa^{\alpha-1}}^{\srg{\frac{2}{\alpha+1}}}-1}{2\,\eps\,\frac{\alpha-1}{\alpha+1}}}^{\srg{\frac{\alpha+1}{2(\alpha-1)}}},\quad \br{\alpha\kappa^{\alpha-1}-1}\,\epsilon>0.$$
For $\eps<0$ and $\alpha\kappa^{\alpha-1}<1$ it is a concave turning point  
with two solutions on the left of that point. For $\eps>0$ and $\alpha\kappa^{\alpha-1}>1$ it is a convex turning point with two solutions on the right of that point, one subsonic and the other supersonic, extending from the common turning point out to the spatial infinity. For $\alpha\kappa^{\alpha-1}=1$ solutions are regular (without shock points) filling all of space: there is a single sonic solution for $\epsilon=0$ given in equation \eqref{eq:regsol}, or a pair of solutions --- supersonic and subsonic for $\eps\ne0$ bifurcating from the previous solution. 

Since the
vector field $[\partial_xF,\partial_yF]$ is everywhere non-zero
except at the stationary point, the qualitative picture sketched in this section should be valid
globally without any new qualitative features.
The example contour lines of the energy surface representing the solutions on the $(x,y)$ plane as well as on the $(x,w)$ plane are sketched in the figures given in section  \ref{sec:spiral} together with
 similar diagrams for spiralling flow.  Three types of solutions  are shown for some example values of parameters $\alpha$ and $\beta$ chosen such that they satisfy, respectively, the conditions: $\gamma<0$ (phase portraits with the elliptic stationary point),  $\gamma=0$, and $\gamma>0$ (phase portraits with the hyperbolic stationary point).     
The observations of the present section are summarised in Table \ref{tab:tableA} where a detailed analysis of the number and kind of sonic points for various combinations of the integration constants $\kappa,\epsilon$ and parameters $\alpha,\beta$ is presented. 

\newcommand{\rpl}[2]{\sout{#1}{\color{red}#2}}
\newcommand{\txt}[1]{\small{\sf{#1}}}
\newcommand{\cola}{1.0cm}
\newcommand{\colb}{1.5cm}
\begin{table}[h!]
\resizebox{\textwidth}{!}{
\centering
\begin{tabular}{@{}|c|c|c|c|c|c|@{}}
\multicolumn{6}{c}{{{ \scshape power-law potentials}}}\\
\hline
$\beta>1$&\txt{accretion}&\txt{specific}&\multicolumn{2}{|c|}{\txt{shock sonic points}}&\multirow{2}{1.5cm}{\txt{the solution domain}}\\
\cline{4-5}
$\alpha>1$&\txt{rate}&\txt{energy}&\txt{\#}&\txt{position}&\empty\\
\hline
%---------------
\multirow{5}{\cola}{$\gamma<0$}&\multirow{5}{\colb}{$\kappa>0$}&
$\eps<\zc<0$&{------}&{------}&{------}\\
\empty&\empty&$\eps=\zc<0{\,\,}%^{(\kappac)}
$&$(1)$&$x_a=\sc=x_b$&$x=\sc$\\
\empty&\empty&$\zc<\eps<0$&$2$&$x_o<x_a<\sc<x_b<\infty$&$x_a<x<x_b$\\
\empty&\empty&$\eps=0$&{$1$}&$x_a=x_o$&$x_a<x<\infty$\\
\empty&\empty&
$\eps>0$&$1$&$0<x_a<x_o$&$x_a<x<\infty$\\
%--------------
\hline
\multirow{7}{\cola}{$\gamma=0$}&\multirow{2}{\colb}{$\kappa^{\alpha-1}>\frac{1}{\alpha}$}&$\eps<0$&{------}&{------}&{------}\\
\empty&\empty&$\eps\geqslant0$&$1$&$x_a=\scc\propto ({\sqrt{\eps}})^{-\frac{\alpha+1}{\alpha-1}}$&$x_a<x<\infty$\\
%--
\cline{2-6}
\empty&\multirow{3}{\colb}{$\kappa^{\alpha-1}=\frac{1}{\alpha}$}&$\eps<0$&{------}&{------}&{------}\\
\empty&\empty&$\eps=0$&$0$&\txt{global sonic accretion}&$y(x)=\alpha^{\frac{1}{1-\alpha}}x^{-\frac{2}{\alpha+1}},\ x>0$\\
\empty&\empty&$\eps>0$&$0$&\txt{sub- or super-sonic accr.}&$0<x<\infty$\\
%--
\cline{2-6}
\empty&\multirow{2}{\colb}{$\kappa^{\alpha-1}<\frac{1}{\alpha}$}&$\eps\leqslant0$&$1$&$x_a=x^{\star\star}\propto ({\sqrt{-\eps}})^{-\frac{\alpha+1}{\alpha-1}}$&$0<x<x_a$\\
\empty&\empty&$\eps>0$&{------}&{------}&{------}\\
\hline
%---------------
\multirow{5}{\cola}{$\gamma>0$}&\multirow{5}{\colb}{$\kappa>0$}&
$\eps<0$&$1$&$0<x_a<x_o$&$0<x<x_a$\\
\empty&\empty&
$\eps=0$&$1$&$x_a=x_o$&$0<x<x_a$\\
\empty&\empty&
$0<\eps<\zc$&$2$&$x_o<x_a<\sc<x_b<\infty$&$0<x<x_a,\quad x_b<x<\infty$\\
\empty&\empty&
$\eps=\zc>0{\,\,}%^{(\kappac)}
$&$(1)$&$\sc$\txt{-centred separatrices}&$0<x\leqslant\sc\leqslant{}x<\infty$\\
\empty&\empty&
$\eps>\zc>0$&$0$&{\txt{sub- or super-sonic accr.}}&$0<x<\infty$\\
\hline
\hline
$\beta>1$&\txt{accretion}&\txt{specific}&\multicolumn{2}{|c|}{\txt{shock sonic points}}&\multirow{2}{1.5cm}{\txt{the solution domain}}\\
\cline{4-5}
$\alpha=1$&\txt{rate}&\txt{energy}&\txt{\#}&\txt{position}&\empty\\
\hline
%---------------
\multirow{3}{\cola}{$\gamma>0$}&\multirow{3}{\colb}{$\kappa>0$}&
$\eps<\zc$&$2$&$x_a<\sc<x_b$&$0<x<x_a,\quad x_b<x<\infty$\\
\empty&\empty&
$\eps=\zc{\,\,}%^{(\kappac)}
$&$(1)$&$\sc$\txt{-centred separatrices}&$0<x\leqslant\sc\leqslant{}x<\infty$\\
\empty&\empty&
$\eps>\zc$&$0$&{\txt{sub- or super-sonic accr.}}&$0<x<\infty$\\
\hline
\multicolumn{6}{c}{}\\
\multicolumn{6}{c}{{\scshape logarithmic potential}}\\
\hline
$\beta=1$&\txt{accretion}&\txt{specific}&\multicolumn{2}{|c|}{\txt{shock sonic points}}&\multirow{2}{1.5cm}{\txt{the solution domain}}\\
\cline{4-5}
$\alpha>1$&\txt{rate}&\txt{energy}&\txt{\#}&\txt{position}&\empty\\
\hline
%---------------
\multirow{3}{\cola}{$\gamma<0$}&\multirow{3}{\colb}{$\kappa>0$}&$\eps<\eps^{\star}$&{------}&{------}&{------}\\
\empty&\empty&$\eps=\eps^{\star}{\,\,}%^{(\kappac)}
$&$1$&$x_a=\sc$&$x=\sc$\\
\empty&\empty&$\eps>\eps^{\star}$&$2$&$0<x_a<\sc<x_b<\infty$&$x_a<x<x_b$\\
%--------------
\hline
\hline
$\beta=1$&\txt{speed}&\txt{specific}&\multicolumn{2}{|c|}{\txt{shock sonic points}}&\multirow{2}{1.5cm}{\txt{the solution domain}}\\
\cline{4-5}
$\alpha=1$&\txt{of sound}&\txt{energy}&\txt{\#}&\txt{position}&\empty\\
\hline
%---------------
\multirow{5}{\cola}{$\gamma=0$}&\multirow{1}{\colb}{$\upsilon<1$}&
$\eps$&$1$&$x_a$&$0<x<x_a$\\
\cline{2-6}
%---------------
\empty&\multirow{3}{\colb}{$\upsilon=1$}&
$\eps<\tilde{\eps}$&{------}&{------}&{------}\\
\empty&\empty&
$\eps=\tilde{\eps}{\,\,}%^{(\kappac)}
$&$0$&\txt{global sonic accr.}&$y(x)=\kappa/x,\quad x>0$\\
\empty&\empty&
$\eps>\tilde{\eps}$&$0$&{\txt{sub- or super-sonic accr.}}&$0<x<\infty$\\
\cline{2-6}
%---------------
\empty&\multirow{1}{\colb}{$\upsilon>1$}&
$\eps$&$1$&$x_a$&$x_a<x<\infty$\\
\hline
\end{tabular}
}
\caption{\label{tab:tableA}
Qualitative properties of cylindrical radial accretion 
in the power-law potential ($\beta>1$) and in the logarithmic potential ($\beta=1$) discussed separately for $\alpha>1$ and $\alpha=1$. The number of shock sonic points is shown in columns marked with \# (if the sonic point is regular the number is parenthesised). Only those of shock sonic points are indicated  which are located at a non-zero finite distance from the centre: $0<x<\infty$). Parameters $x_a$ and $x_b$ are the roots of the equation $\mathcal{Z}(x)=\epsilon$ and represent the positions of two shock sonic points (if there is only a single solution then it is denoted by $x_a$). Parameter $x_o$ is the root of the equation $\mathcal{Z}(x)=0$.  Parameters $\sc$, $\scc$ and $\zc$ are defined in the respective sections of the text.
The solution domain is the region where solutions exist.}
\end{table}

\subsection{\label{sec:alpha1powlaw} Polytropic exponent $\alpha=1$ ($\gamma=2(\beta-1)>0$)}

\noindent
According to equations \eqref{eq:Upotential},
and \eqref{eq:intsurf1} in this case the first integrals can be recast in a single unifying form
\begin{equation}\label{eq:FBetaAlpha1}F(x,y)\equiv\frac{1}{2}\frac{\kappa^2}{x^2y^2}-\frac{1}{\beta-1}\frac{1}{x^{\beta-1}}+\upsilon^2\ln{y}, \qquad x>0,\quad y>0.
\end{equation}
Finding the level lines of constant specific energy $\eps$ for the above form of $F(x,y)$ with the logarithm leads to a more general problem (characteristic of various accretion problems with $\alpha=1$) which is to find a solution $y(x)$ to an equation with the following general structure
\begin{equation}\label{eq:GenAccrEq}
y^{-m}\,\mathcal{K}(x)+\ln{y}-\mathcal{E}(x)=0.\end{equation}
In this case, the solution $y(x)$ can be found in an exact form expressible in terms of a transcendental analytic function (the same concerns the case with $x,y$ interchanged, which we consider later,  when we obtain a reversed solution $x(y)$).
 By substituting $\ln{y}=\frac{1}{m}\,\omega+\mathcal{E}(x)$ 
 we obtain an equation $\omega\,\expp{\omega}=-m\,\mathcal{K}(x)\,\expp{-m\,\mathcal{E}(x)}$ for a new unknown $\omega$. Knowing that $W(\omega\,\expp{\omega})\equiv\omega$ by definition of the (multivalued) Lambert $W$ function, we obtain
\begin{equation}\label{eq:GenAccrEqSol}y(x)=
\exp\sq{\mathcal{E}(x)+\frac{1}{m}\,W\br{-m\,\mathcal{K}(x)\,\expp{-m\,\mathcal{E}(x)}}},
\end{equation}
which solves equation \eqref{eq:GenAccrEq}.
The Lambert $W$ function has found many applications in the literature although its presence often goes unrecognised, as noticed by \citet{corless1996} who collect many available results on that function.
 On the real line $\xi$, $W(\xi)$  attains real values for $\xi\geqslant-\frac{1}{e}$. It is a negative double-valued function for $-\frac{1}{e}<\xi<0$ in the region of interest here. The principal branch  satisfying $W(x)\geqslant-1$ is denoted by $W_0(x)$, and the branch satisfying $W(x)\leqslant-1$ by $W_{-1}(\xi)$. For $\xi>0$ not of interest here, $W_{-1}(\xi)$ is positive. 
On identifying $\mathcal{K}(x)=\frac{1}{2}\frac{\kappa^2}{x^2\upsilon^2}$ and $\mathcal{E}(x)=\frac{1}{\upsilon^2}\br{\eps+\frac{1}{\beta-1}\frac{1}{x^{\beta-1}}}$  we find that 
$$y(x)=
\exp\sq{\frac{1}{\upsilon^2}\br{\epsilon+\frac{x^{1-\beta}}{\beta-1}}+\frac{1}{2} {W}\br{-\frac{\kappa^2}{x^2\upsilon^2}\exp\sq{-\frac{2}{\upsilon^2}\br{\eps+\frac{x^{1-\beta}}{\beta-1}}}}}
$$ in the particular case of $F(x,y)$ defined in equation \eqref{eq:FBetaAlpha1}. In order to obtain complete level lines we must use both branches of function $W$ in this expression for $y(x)$. 

The analysis which follows is quite analogous to that of the previous section \ref{sec1}, though we will limit ourselves to presenting the results only.
With the Hamiltonian \eqref{eq:FBetaAlpha1} there is a single stationary point of the energy surface
$$\xc={\lrg{\upsilon^{-\frac{2}{\beta-1}}}},\qquad \yc=\lrg{\kappa\,\upsilon^{{{\frac{3-\beta}{\beta-1}}}}},\qquad \zc=\frac{\upsilon^2}{2}\ln{\br{\kappa^2\br{\frac{\upsilon^2}{e}}^{\srg{\frac{3-\beta}{\beta-1}}}}}.$$
 The corresponding critical accretion rate  $\kappac$ for solutions with given specific energy $\eps$ is 
$$\kappac=\lrg{\upsilon^{\frac{\beta-3}{\beta-1}}}\exp\br{\frac{\eps}{\upsilon^2}+\frac{3-\beta}{2(\beta-1)}}.$$
The stationary point is hyperbolic, because the Hessian determinant is negative at that point 
$$
\left.\det{[\partial^2_{ij}F(x,y)]}\right|_{{x=\xc}\atop{y=\yc}}= -2\br{\beta-1}\frac{\upsilon^6}{\kappa^2}<0.$$ 

\noindent
As previously, the shock curve is defined 
as a subset of the energy surface determined by the constraint $\partial_{y} F(x,y)=0$ for $x\neq\xc$, hence, parameterized with $x$, that curve reads
$$y=\mathcal{Y}(x)=\frac{\kappa}{\upsilon\,x},\qquad \mathcal{Z}(x)\equiv F(x,Y(x))=\frac{1}{2}\upsilon^2+\upsilon^2\ln\br{\frac{\kappa}{\upsilon\,x}}
-\frac{1}{\beta-1}\frac{1}{x^{\beta-1}}.$$
The stationary point is the global maximum on the shock curve as $x\to\xc$: the statement follows from the fact that $
\mathcal{Z}'(\xc)=0$ and $\mathcal{Z}''(\xc)=-\br{\beta-1}\upsilon^{\frac{2(\beta+1)}{\beta-1}}<0$ (note also that $\mathcal{Z}(x)\to-\infty$ as $x\to0$ or $x\to+\infty$).
The turning points of integral curves represented as level lines $x(y)$ (for which $x'(y)=0$ on the shock curve) are concave for $x<\xc$ and convex for $x>\xc$ as seen from the expression for $x''(y)$ evaluated on the shock curve
$$x''(y)|_{y=\mathcal{Y}(x)}=\frac{2\upsilon^4\,x^{2+\beta}}{\kappa^2\br{\br{\frac{x}{\xc}}^{\beta-1}-1}}.$$  
Projected onto the $(x,y)$ plane, the shock curve overlaps with the sonic curve $y(x)=\frac{\kappa}{\upsilon\,x}$ defined as one for which $w(x,y)=\upsilon$. 

In power-law potentials we have $\gamma>0$ if $\alpha=1$. It turns out that  the shape of the  level lines on this plane for $\alpha=1$ is qualitatively the same as for previously discussed solutions with  $\alpha>1$ and $\beta$ such that $\gamma>0$ too (provided $\epsilon<0$ for the latter solutions). The number of shock sonic points  and the solution domain in function of energy for $\alpha=1$ can be seen in Table \ref{tab:tableA} and compared with the $\alpha>1$ case. 
A variety of position-density and position-velocity phase diagrams in the power-law potential will be presented in section \ref{sec:spiral} where the diagrams for the radial accretion is compared with the corresponding diagrams for the spiralling accretion. 

\section{Application to Bondi model} 

The same analysis as we have carried out for the radial cylindrical accretion can be
 applied to Bondi accretion (spherically symmetric radial accretion in the Newtonian potential). We also use the same convention for units remembering that now $x$ represents the radial variable in spherical coordinates.
The Bondi accretion can be described by the Hamiltonian function
$$F(x,y)=\frac{1}{2}\frac{\kappa^2}{x^4y^2}+\frac{\alpha}{\alpha-1}y^{\alpha-1}-\frac{1}{x},\qquad{}x>0,\quad{}
y>0.$$
The important change in comparison with the cylindrical accretion is in the kinetic term. It is due to different geometrical interpretation of the radial variable $x$ (the  continuity integral now reads $w(x,y)x^2y=\kappa$). The Hessian determinant evaluated at the stationary point $(\xc,\yc)$ of the integral distinguishes $\alpha=5/3$ as the critical value  of the polytropic exponent:
$$\left.\det{[\partial^2_{ij}F(x,y)]}\right|_{{x=\xc}\atop{y=\yc}}=-(5-3\alpha)\frac{\br{\alpha(4\kappa)^{\alpha-1}}^{\srg{\frac{2}{5-3\alpha}}}}{{\xc}^3{\yc}^2}.$$
{The resulting specific energy at the stationary point for Bondi accretion is $\zc=\frac{5-3\alpha}{4(\alpha-1)\xc}$ with $\xc={\left( {2}^
      \frac{\alpha  + 1}{2}\,\alpha \,
     {\kappa }^{\alpha  - 1}
     \right) }^
  {-\ssrg{\frac{2}{{5-3\,\alpha }}}}$, $\yc=\br{\frac{\kappa^2}{\alpha\,(\xc)^4}}^{\frac{1}{\alpha+1}}$. The corresponding critical accretion rate  $\kappac$ for solutions with given specific energy $\eps$ is 
\newcommand{\flatfrac}[2]{\br{#2}^{-1}{#1}}  
  $$\kappac  = {\left( \flatfrac{{\left(
           \frac{4\,\epsilon \,
            \left( \alpha  - 1
            \right) }{5 - 
            3\,\alpha } \right) }^
        {\frac{5 - 3\,\alpha }{2}}}
        {\alpha \,
        2^{\frac{\alpha  + 1}{2}}}
      \right) }^
   {\frac{1}{\alpha  - 1}}$$.} 
   For $\alpha<5/3$ the stationary point is hyperbolic and it is also the global maximum on the shock sonic curve $$\mathcal{Z}(x)=\frac{1}{x}\,\left( \frac{1}{2}\,
     \frac{\alpha  + 1}
      {\alpha  - 1}\,
     {\left( \alpha \,
         {\kappa }^{\alpha  - 1}
         \right) }^
      {\frac{2}{\alpha+1 }}\,
     \lrg{x^{\frac{5 - 3\,\alpha }
        {\alpha +1}}} - 1 \right).$$  For $\alpha>5/3$ the stationary point is elliptic and it is the global minimum of the energy surface. 
                In both cases, below and above $\alpha=5/3$, the structure of solutions 
on the phase plane $(x,y)$ is qualitatively the same as for cylindrical radial accretion in power-law potentials with $\gamma>0$ or $\gamma<0$, respectively.
The case of critical exponent $\alpha=5/3$ corresponds to critical parameter $\gamma=0$ of cylindrical radial accretion. Again, the structure of solutions is qualitatively the same -- for Bondi accretion there is a limiting value $\kappa_{5/3}=\frac{3}{20}\sqrt{3/5}$ delineating  solutions with single sonic shocks (present for $\kappa\neq\kappa_{5/3}$)
 from global accretion solutions with no shocks (for $\kappa=\kappa_{5/3}$). The limiting value $\kappa_{5/3}$  for Bondi accretion can be obtained by requiring that $\mathcal{Z}'(x)\equiv0$. 

 For $\alpha=1$, the Hamiltonian
attains the form
\begin{equation}\label{eq:bondiF}F(x,y)=\frac{1}{2}\frac{\kappa^2}{x^4y^2}+\upsilon^2\ln{y}-\frac{1}{x}=\eps,\qquad{}x>0,\quad{}
y>0.\end{equation}
Again, similarly to the cylindrical case with $\alpha=1$, the above equation possess the general form of equation \eqref{eq:GenAccrEq}, therefore the solution $y(x)$ can be found in an exact form (note, the formal difference in the kinetic and potential term when compared with the cylindrical counterpart given in equation \eqref{eq:FBetaAlpha1}). On substituting $y(x)=\exp{\br{\frac{1}{2}\omega(x)+\frac{1}{\upsilon^2}\br{\eps+\frac{1}{x}}}}$ in equation \eqref{eq:bondiF}, it follows that $\omega\,\expp{\omega}=-\frac{\kappa^2}{\upsilon^2x^4}\exp\br{-\frac{2}{\upsilon^2}\br{\eps+\frac{1}{x}}}$. Since $W(\omega\,\expp{\omega})\equiv\omega$ by the defining property of the Lambert $W$ function, we obtain the following radial profile: 
$$y(x)=
\exp\sq{\frac{1}{\upsilon^2}\br{\epsilon+\frac{1}{x}}+\frac{1}{2}W\br{-\frac{\kappa^2}{x^4\upsilon^2}\exp\sq{-\frac{2}{\upsilon^2}\br{\eps+\frac{1}{x}}}}}.
$$ It depends on two integration constants $\epsilon$ and $\kappa$ (the formal distinction between accretion and winds lies in the sign of $\kappa$ which is important in the sign of the inflow velocity: $w(x)=\frac{\kappa}{x^2 y(x)}$ -- in our notation $\kappa<0$ for winds and $\kappa>0$ for accretion).
All information about the classical Bondi accretion or Parker's winds with the polytropic exponent $\alpha=1$ is contained in the above formula. The first application of the Lambert function in the context of classical Bondi model was given by \citet{cranmer2004} who solved exactly the 
 classical Parker's solar wind problem for an isothermal plasma. For other spherical potentials in the context of Bondi model  see \citep{Ciotti_2017}.
 
\section{\label{sec:spiral}Spiralling accretion in the power-law potential \  ($\alpha\geqslant1, \ \beta>1$)}

The centrifugal potential present in the general form of the Hamiltonian $F(x,y)$ in equation
\eqref{eq:intsurf1} accounts for the dynamical effect of the conservation of specific angular momentum. This term plays a substantial role close to the centre.  As a result,  the phase diagram pictures in this region may be deformed to such extent so that new qualitative features (like additional shock points) may appear not observed in the case of purely radial accretion.

\subsection{ Spiralling accretion for $\alpha>1$ and $\beta>1$}

\begin{figure}[h]
\begin{center}
\centering
\includegraphics[width=0.3\textwidth]{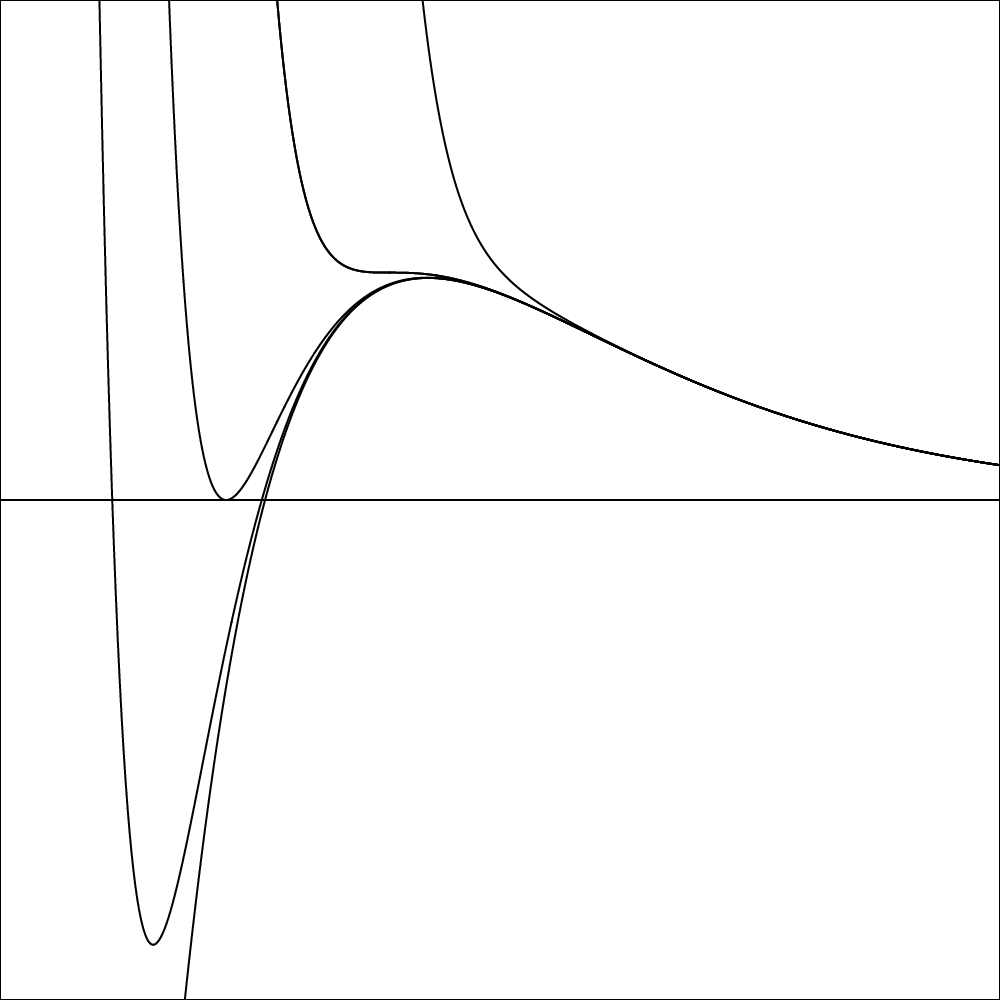}
\caption{\label{fig:03} The level function $\mathcal{Z}(x)$ of the shock curve sketched for a fixed $\gamma$ ($0<\gamma<4$) and several values of $\lambda$ about its critical value $\lsfl$ (for which $\mathcal{Z}(x)$  possess a stationary inflection point):  $\lambda>\lsfl$ (the curve with no extrema), $\lambda=\lsfl$ (the curve with the stationary inflection point), $\lambda<\lsfl$ (two curves with a minimum and a maximum), and $\lambda=0$ -- the case of radial accretion (the envelope curve of remaining curves). } 
\end{center}
\end{figure}

For $\lambda\neq0$ the corresponding unifying integral defining the energy surface $F(x,y)$ is given by
\begin{equation}
F(x,y)=\frac{1}{2}\frac{\kappa^2}{x^2y^2} +\frac{1}{2}\frac{\lambda^2}{x^2}-\frac{1}{\beta-1}\frac{1}
{x^{\beta-1}}+\frac{\alpha}{\alpha-1}\,y^{\alpha-1},\qquad{}x>0,\quad{}
y>0.\label{eq:intsurf2}\end{equation}
The shock curve is accordingly also modified by the centrifugal potential 
\begin{equation*}\label{eq:shockcurvelambda}
\left.\begin{array}{l} 
	\!\! \rplcd{X(x)=x,\quad
	Y(x)}{y=\mathcal{Y}(x)},
	%	\qquad 
	%\\
	\qquad \mathcal{Z}(x)\equiv F(x,\mathcal{Y}(x))=\lrg{\frac{1}{2}\frac{\lambda^2}{x^2}}+
	\left\{\begin{array}{l}
\lrg{\frac{1}{2}\frac{\alpha+1}{\alpha-1}\br{\alpha\kappa^{\alpha-1}}^{
\frac{2}{\alpha+1}}x^{-2\frac{\alpha-1}{\alpha+1}}-\frac{1}{\beta-1}
x^{-(\beta-1)}}\quad\mathrm{for}\ \gamma\ne0,\\
\lrg{\frac{1}{2}\frac{\alpha+1}{\alpha-1}\sq{\br{\alpha\,
\kappa^{\alpha-1}}^{\frac{2}{\alpha+1}}-1}
x^{-2\frac{\alpha-1}{\alpha+1}}}\quad
\mathrm{for}\ \gamma=0\,\wedge\,\alpha\,
\kappa^{\alpha-1}\neq1.
 			\end{array}\right.
	\end{array}\right.
\end{equation*}
The function $\mathcal{Z}(x)$ is sketched in Figure \ref{fig:03} for several values of the parameter $\lambda$. The number of shock points is equal to the number of separate real roots of the equation $\mathcal{Z}(x)=\eps$.

\begin{figure}[h!]
\centering
\includegraphics[width=0.9\textwidth]{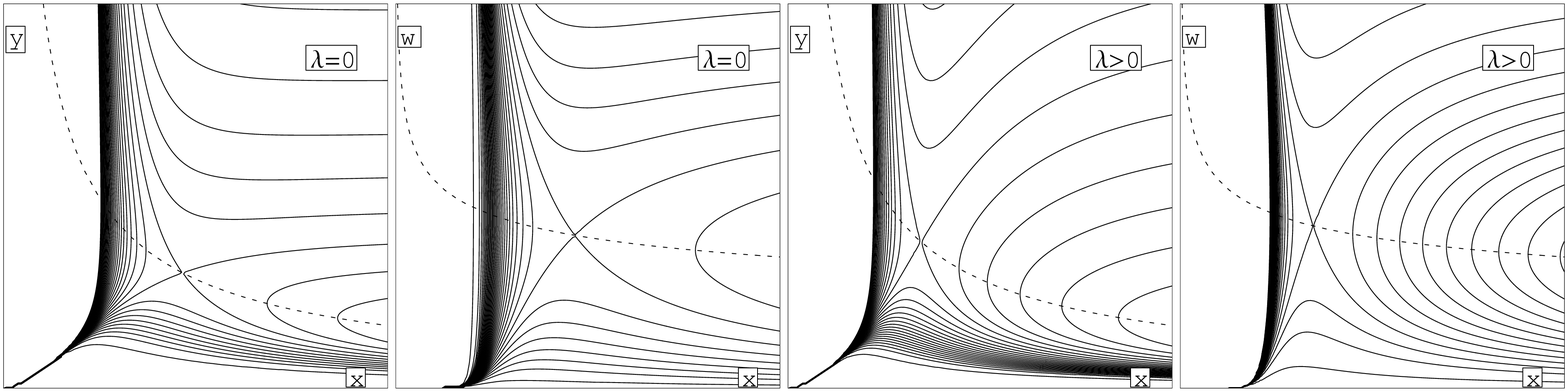}
\caption{\label{fig:gammabeyondfour}
 $\alpha>1$, $\beta>3$ ($\gamma>4$) $\diamond$ Radial and spiralling cylindrical accretion of polytropic matter in the
power law potential illustrated with example level lines of the energy surface, shown in the position-density $(x,y)$ and in the position-velocity $(x,w)$ phase planes. 
{\it 1st}: density function for radial accretion ($\lambda=0$), 
{\it 2nd}: radial velocity for radial accretion ($\lambda=0$), 
{\it 3rd}: density function for spiralling accretion ($\lambda>0$), 
{\it 4th}: radial velocity for spiralling accretion ($\lambda>0$).
 The figures were prepared by assuming $\alpha<\beta$. The shock curve is shown with the dashed line (also in all other phase diagram figures that follow).
}
\includegraphics[width=0.9\textwidth]{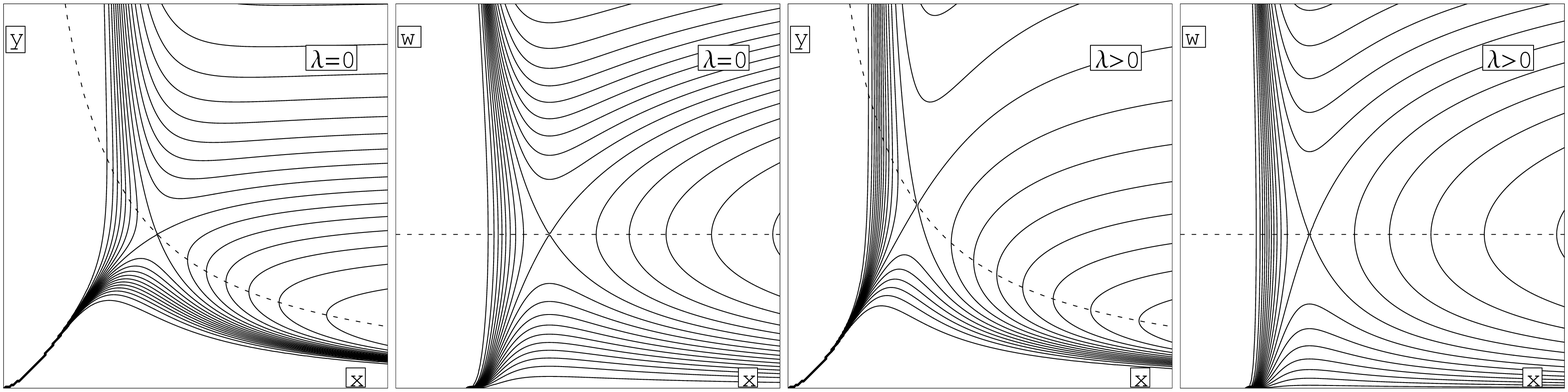}
\caption{\label{fig:gammabeyondfourisot} $\alpha=1$, $\beta>3$ ($\gamma>4$) $\diamond$ Radial and spiralling cylindrical accretion of isothermal matter in the
power law potential. The diagrams correspond with the same parameters to respective diagrams in Figure \ref{fig:gammabeyondfour}. }
\end{figure}

\noindent
{\bf A)} {\it Spiralling accretion for $\gamma>4$ ($\alpha>1$, $\beta>3$)}.
The function $\mathcal{Z}(x)$ 
starts from $-\infty$ ($\mathcal{Z}(x)\searrow-\infty$ as $x\searrow0$)
and then it is increasing until 
it attains a global maximum which is positive (this is the only extremum and one infers from equations \eqref{eq:id2} that $\mathcal{Z}(x)>0$ and $\mathcal{Z}''(x)<0$  at the extremum). Then the function is decreasing and $\mathcal{Z}(x)\searrow0$ as $x\nearrow+\infty$.  
The shock curve is thus qualitatively the same as the one for the radial flow ($\lambda=0$) with $\gamma>0$. 
The only stationary point of the energy surface coinciding with the extremum is hyperbolic. 
 From this we conclude, that for $\gamma>4$ the phase diagram  of the spiralling flow on the $(x,y)$ or $(x,w)$ phase planes  will be  qualitatively similar to the corresponding phase diagrams of the radial flow. This can be seen in Figure 
\ref{fig:gammabeyondfour}. As shown in section \ref{sec:alpha1powlaw} and later in section \ref{sec:saisot}, analogous solutions in the isothermal case ($\alpha=1$) can be expressed in analytical form. For comparison, their phase portraits for $\gamma>4$ are shown in Figure \ref{fig:gammabeyondfourisot}. As can be seen, there is no qualitative difference between phase diagrams in both figures, which suggests that the limit $\alpha\to1$ is continuous.

\medskip
\noindent
{\bf B)} {\it Spiralling accretion for $\gamma=4$ ($\alpha>1$, $\beta=3$)}.
For $\gamma=4$  the exact solution can be found, however in the reversed form
\begin{equation}\label{eq:solx}
x(y)=\frac{1}{y}\,{\sqrt{\frac{{\kappa }^2 + 
        \left( {\lambda }^2 - 1 \right) \,y^2}{2\,
        \left( \epsilon  - 
          \frac{\alpha }{\alpha  - 1}\,y^{\alpha  - 1}
          \right) }}},\qquad w(y)=\frac{\kappa}{y\,x(y)}.\end{equation}
The constant density solution not included above reads (with a parameter $\phi$ enumerating admissible $\lambda$)
$$u(x)=\frac{\sin{\phi}}{x},\qquad w(x)=\frac{\cos{\phi}}{x}, \qquad y(x)=\br{\frac{\alpha-1}{\alpha}\eps}^{\frac{1}{\alpha-1}} \qquad \mathrm{at}\quad
\lambda=\sin{\phi},\quad \kappa=\br{\frac{\alpha-1}{\alpha}\eps}^{\frac{1}{\alpha-1}}\cos{\phi}>0.$$         
        In order to describe the  solution \eqref{eq:solx} qualitatively, we must consider three cases: 
        
{\it i)} $\lambda^2<1$.
The function $\mathcal{Z}(x)$ 
starts from $-\infty$ ($\mathcal{Z}(x)\searrow-\infty$ as $x\searrow0$)
and then it is increasing, attains its $0$ at $x={\left( \frac{\alpha  - 1}{\alpha  + 1}\,
      \left( 1 - {\lambda }^2 \right)  \right) }^
   {\frac{\alpha+1 }{4}}\,
  \frac{1}{{\sqrt{\alpha \,{\kappa }^{\alpha  - 1}}}}$ then it still grows until 
it attains a global maximum $\frac{\alpha \,{\kappa }^{\alpha  - 1}}{\alpha  - 1}\,
  {\left( 1 - {\lambda }^2 \right) }^
   {\frac{1 - \alpha }{2}}>0$ at $x=\frac{{\left( 1 - {\lambda }^2 \right) }^
    {\frac{\alpha+1 }{4}}}{{\sqrt{\alpha \,
       {\kappa }^{\alpha  - 1}}}}$. Then the function is decreasing and $\mathcal{Z}(x)\searrow0$ as $x\nearrow+\infty$.  The shock curve is qualitatively the same as the one for the radial flow ($\lambda=0$) with $\gamma>0$. 
The stationary point at the maximum is hyperbolic. 
 From this we conclude, that for $\gamma=4$ the phase diagram  of the spiralling flow on the $(x,y)$ plane is  qualitatively similar to the phase diagram of the radial flow with $\gamma>0$.

 {\it ii)} $\lambda^2=1$. In this case the potential term is cancelled by the centrifugal term. The function $\mathcal{Z}(x)$ 
starts from $+\infty$ ($\mathcal{Z}(x)\nearrow+\infty$ as $x\searrow0$)
and then it is monotonically decreasing until it attains $0$ at $x=+\infty$.
In this case the function $F(x,y)$ is positive and we must set $\eps>0$. The level lines $F(x,y)=\eps>0$ have left-sided turning points on the shock curve  because the function $x(y)$ is convex at the intersection point at $x_{\eps}=\kappa \,{\alpha }^{\frac{1}{\alpha  - 1}}\,
  {\left( \frac{1}{2\,\epsilon }\,
      \frac{\alpha  + 1}{\alpha  - 1} \right) }^
   {\frac{1}{2}\,\frac{\alpha  + 1}{\alpha  - 1}}$, that is $x''\left|_{y=\mathcal{Y}(x_{\eps})}\right.=\kappa \,\left( \alpha+1  \right) \,
  {\alpha }^{\frac{3}{\alpha  - 1}}\,
  {\left( \frac{\alpha  + 1}
      {2\,\epsilon \,\left( \alpha  - 1 \right) }
      \right) }^
   {\frac{\alpha+5 }{2\,\left( \alpha  - 1 \right) }}$ is positive.   
  The solution $y(x)$ corresponding to equation \eqref{eq:solx} with $\lambda^2=1$ consists of two branches, the lower branch and the upper branch, respectively, with $y$ lower or greater than $y_{\eps}={\left( \frac{2\,\epsilon \,
       \left( \alpha  - 1 \right) }{\alpha \,
       \left( \alpha  + 1 \right) } \right) }^
  {\frac{1}{\alpha  - 1}}$.

{\it iii)} $\lambda^2>1$.  The function $\mathcal{Z}(x)$ 
starts from $+\infty$ ($\mathcal{Z}(x)\nearrow+\infty$ as $x\searrow0$)
and then it is monotonically decreasing until it attains $0$ at $x=+\infty$.
The function $F(x,y)$ is positive. The level lines $F(x,y)=\eps>0$ have left turning points on the shock curve  because the function $x(y)$ is convex at the intersection point. 

\begin{figure}[h!]
\centering
\includegraphics[width=0.9\textwidth]{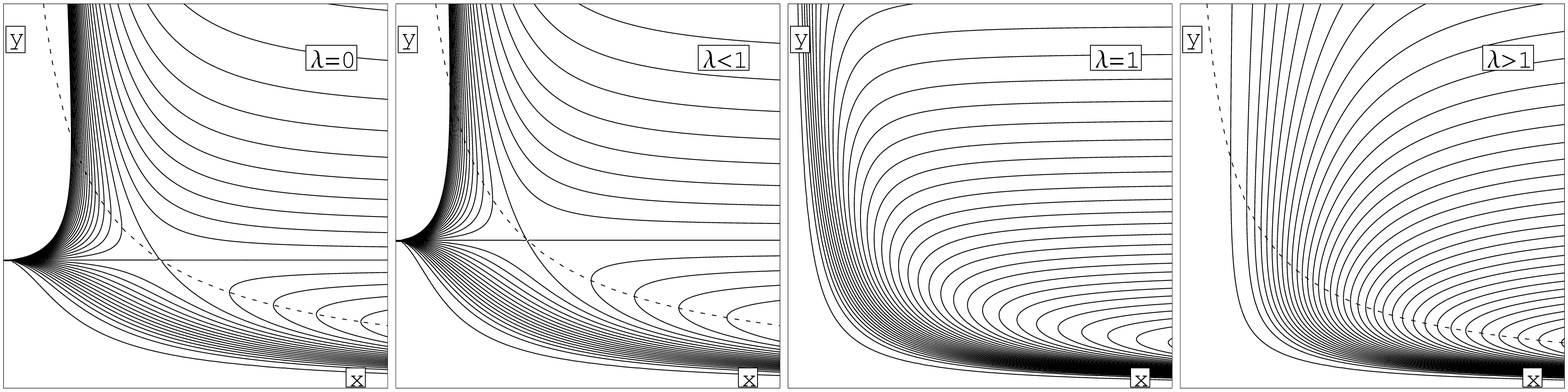}
\includegraphics[width=0.9\textwidth]{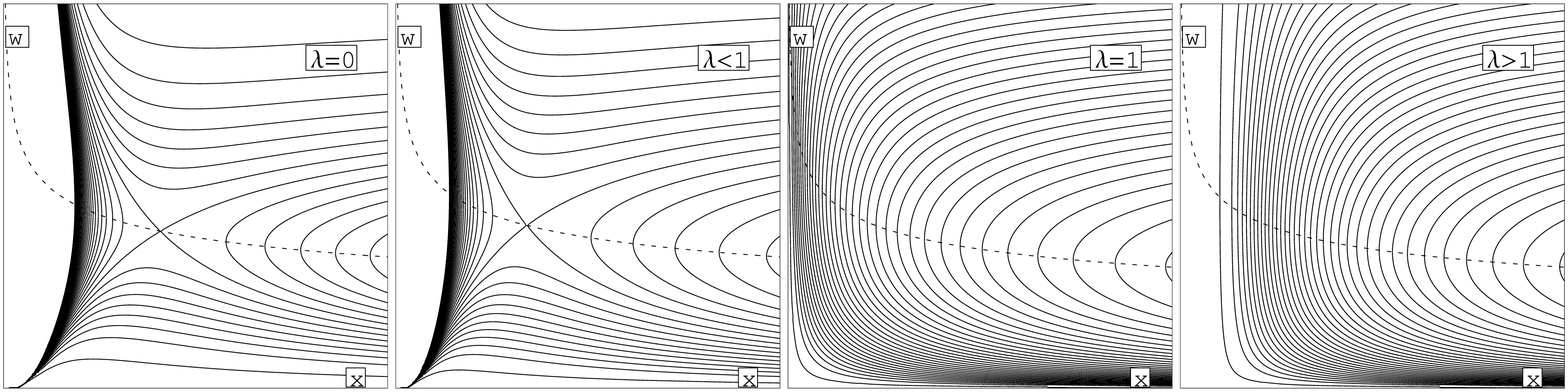}
\caption{\label{fig:gammafour} $\alpha>1$, $\beta=3$ ($\gamma=4$) $\diamond$
Radial and spiralling cylindrical accretion of polytropic matter in the
power law potential illustrated with example level lines of the energy surface, shown for various $\lambda$ in the position-density $(x,y)$ phase plane  {\it [upper row]}  and in the position-velocity $(x,w)$ phase plane {\it [bottom row]}: [{\it 1st column}] radial accretion ($\lambda=0$), [{\it columns 2nd,3rd,4th}] spiralling accretion for $0<\lambda<1$, $\lambda=1$  and $\lambda>1$, respectively.
The figures were prepared by assuming $\alpha<\beta$}
% $\alpha=3/2$, $\beta=3$, $\kappa=1$, $\lambda=0,1/2,1,2$.}
\includegraphics[width=0.9\textwidth]{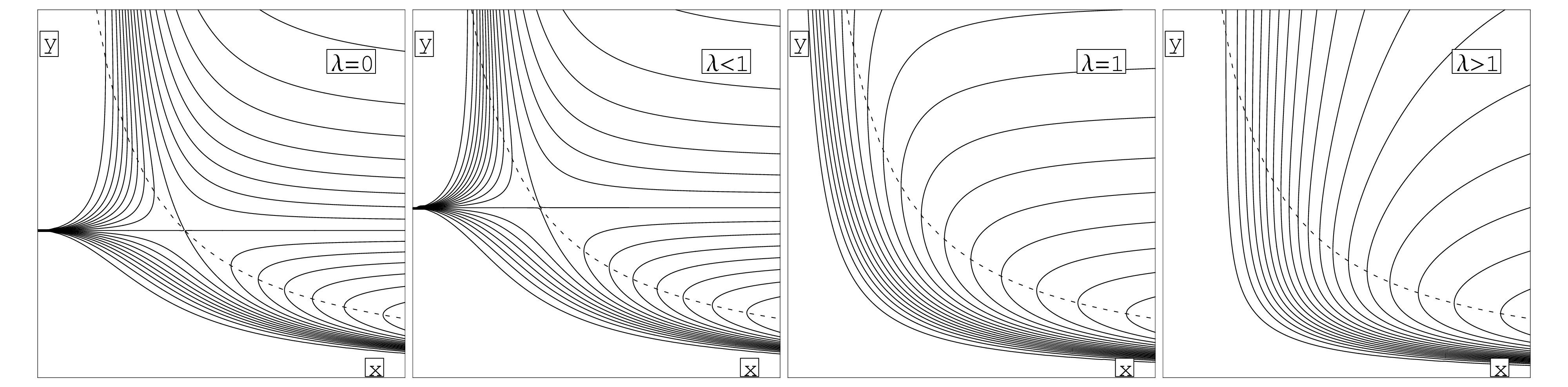}
\includegraphics[width=0.9\textwidth]{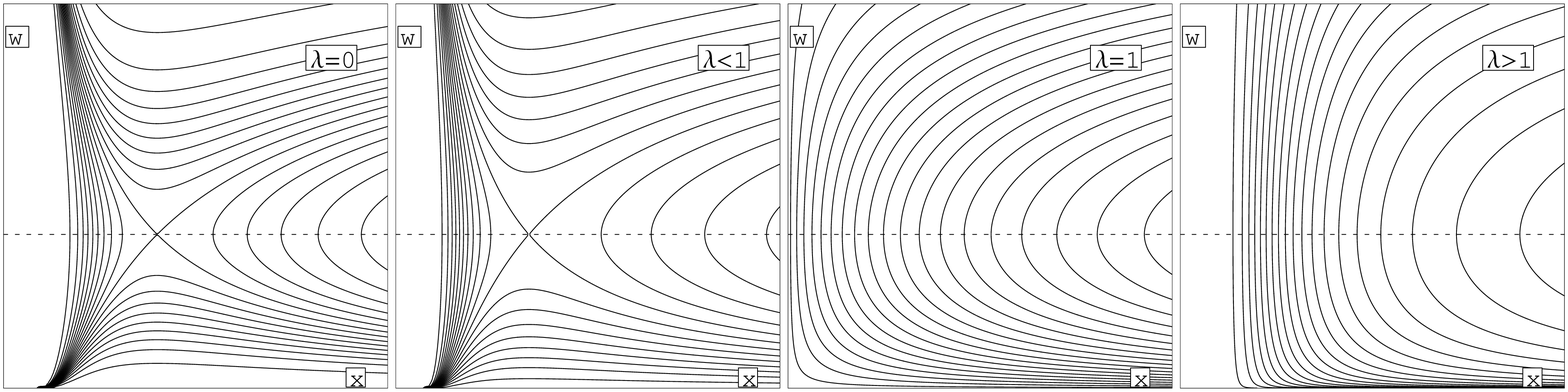}
\caption{\label{fig:gammafouriso} $\alpha=1$, $\beta=3$ ($\gamma=4$) $\diamond$
Radial and spiralling cylindrical accretion of isothermal matter in the
power law potential. The diagrams correspond with  the same parameters (for some $\upsilon>0$) to respective diagrams in Figure \ref{fig:gammafour}.}
\end{figure}

The phase portraits for $\gamma=4$ and $\alpha>1$ are shown in Figure \ref{fig:gammafour} for various parameters $\lambda$ distinguished by the form of solution in equation \ref{eq:solx}, including also the purely radial flow. For comparison, in Figure \ref{fig:gammafouriso} are shown corresponding phase portraits of the limiting isothermal solution ($\alpha=1$) for $\gamma=4$, which also can be found in exact form, as shown later in section \ref{sec:saisot}. There is no qualitative difference
between phase diagrams in both figures, similarly as it was for $\gamma>4$.

\medskip

\begin{figure}[h!]
\centering
\includegraphics[width=0.9\textwidth]{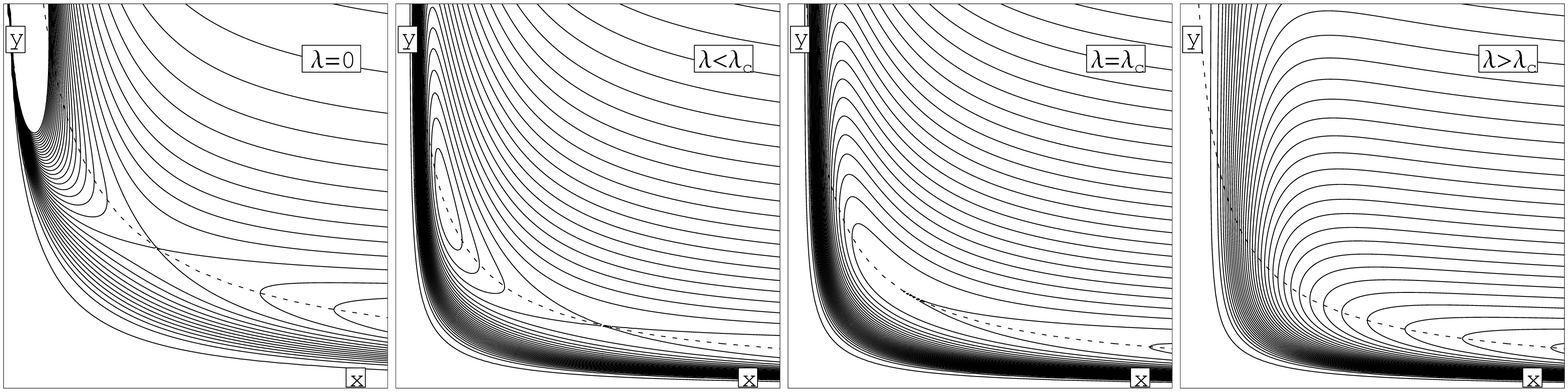}
\includegraphics[width=0.9\textwidth]{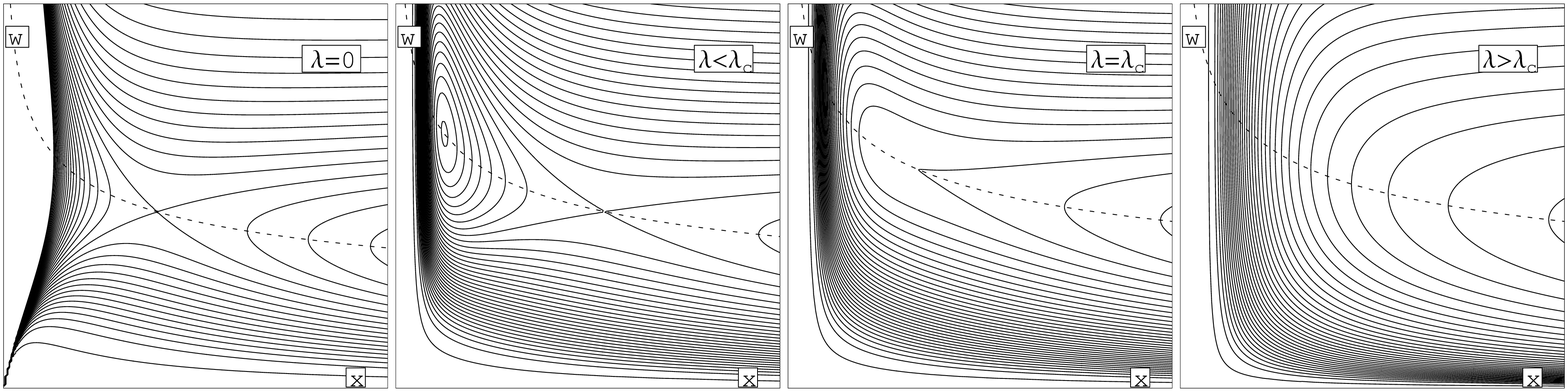}
\caption{\label{fig:gammaphysunitalphgt}
$\alpha>1$, $\frac{3\alpha-1}{\alpha+1}<\beta<3$ ($0<\gamma<4$) $\diamond$
Radial and spiralling cylindrical accretion of polytropic matter in the
power law potential. Example level lines of the energy surface, shown for various $\lambda$ about the critical value $\lsfl$ in the position-density $(x,y)$ plane  {\it [upper row]} and in the position-velocity $(x,w)$ plane {\it [bottom row]}: [{\it 1st column}] 
 radial accretion ($\lambda=0$), [{\it columns 2nd,3rd,4th}] spiralling accretion 
for $0<\lambda<\lsfl$, $\lambda=\lsfl$  and $\lambda>\lsfl$, respectively.  
The figures were prepared by assuming $\alpha<\beta$ (for $\alpha>\beta$ and $\lambda=0$ the radial velocity  would diverge at $x=0$ and for $\alpha=\beta$ attain a finite value at $x=0$ (not shown)). }
%$\alpha=5/3$, $\gamma=2$, $\kappa=1$, $\lambda=0,(2\locl+\lsfl)/3,\lsfl,%3\lsfl/2$. . }
\includegraphics[width=0.9\textwidth]{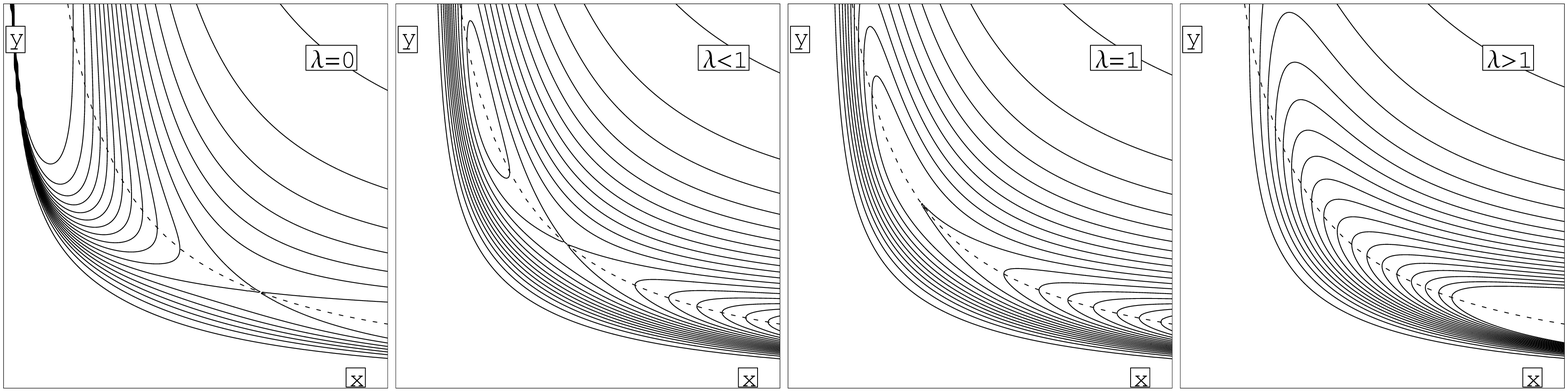}
\includegraphics[width=0.9\textwidth]{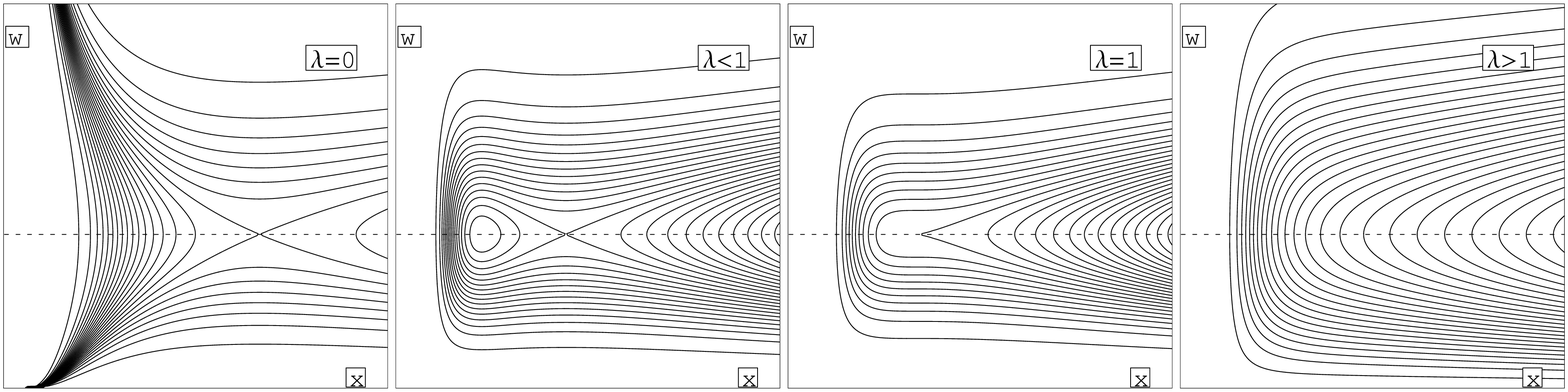}
\caption{\label{fig:gammaphysunitalph}
$\alpha=1$, $1<\beta<3$ ($0<\gamma<4$) $\diamond$
Radial and spiralling cylindrical accretion of isothermal matter in the
power law potential. The diagrams correspond to respective diagrams in Figure \ref{fig:gammaphysunitalphgt}.}
\end{figure}

\noindent
{\bf C)} {\it Spiralling accretion for $0<\gamma<4$ ($\alpha>1$)}.       
For $\lambda\neq0$, the situation is substantially changed close to the center in comparison with the radial flow. In this case the function $\mathcal{Z}(x)$ is shown in Figure \ref{fig:03} for some example value of $\gamma$. Now $\mathcal{Z}(x)\nearrow+\infty$ as $x\searrow0$, then $\mathcal{Z}(x)$ is rapidly decreasing and attains some local minimum which can be negative 
or positive depending on $\lambda$. The minimum will be zero at some $x$ for which both $\mathcal{Z}=0$ and $\mathcal{Z}'(x)=0$. We call such point a stationary osculating point. By solving these conditions for $x$ and $\lambda$ we obtain
$$\xocl={\left( \frac{2\,\left( \alpha  - 1 \right) }
      {\left( \alpha  + 1 \right) \,\left( \beta  - 1 \right) }\,
     \left( 1 - \frac{\gamma }{4} \right)  \right) }^
  {\frac{\alpha  + 1}{\gamma }}\xc,  \qquad
\locl^2=\frac{\gamma }{2\,\left( \beta  - 1 \right) }\,
  {\left( \frac{\frac{2\,\left( \alpha  - 1 \right) }
         {\left( \alpha  + 1 \right) \,
           \left( \beta  - 1 \right) }\,
        \left( 1 - \frac{\gamma }{4} \right) }{{\left( \alpha \,
           {\kappa }^{\alpha  - 1} \right) }^
        {\frac{2}{\alpha  + 1}}} \right) }^
   {\frac{4}{\gamma } - 1},\qquad  0<\gamma<4,
  \quad \xc=\frac{1}{{\left( \alpha \,{\kappa }^{\alpha  - 1} \right) }^
    {\frac{2}{\gamma }}}.$$
    $$\det{[\partial^2_{ij}F(\xocl,\mathcal{Y}(\xocl))]}>0,\qquad
\partial^2_{xx}F(\xocl,\mathcal{Y}(\xocl))=     \frac{2\,\frac{\beta  - 1}{\alpha  - 1}\,
    {\left( \alpha \,{\kappa }^{\alpha  - 1} \right) }^
     {\frac{2\,\left( 1 + \beta  \right) }{\gamma }}}{{\left( \frac{2\,
          \left( \alpha  - 1 \right) }{\left( \alpha  + 1 \right) \,
          \left( \beta  - 1 \right) }\,\left( 1 - \frac{\gamma }{4} \right)  \right)
       }^{\frac{4\,\alpha }{\gamma }}}.$$
Because the determinant and trace of the Hessian matrix are both positive at
the stationary osculating point, the point is also a local minimum of the energy surface $F(x,y)$. As $x$ gets  higher,
 $\mathcal{Z}(x)$ is increasing and attains a global positive maximum, then  $\mathcal{Z}(x)$ is decreasing and $\mathcal{Z}(x)\searrow0$ as $x\nearrow+\infty$. This situation is changed further for $\lambda>\lsfl$ for some $\lsfl$, such that in the limit $\lambda=\lsfl$  both extrema
coalesce and become a single stationary inflection point
of $\mathcal{Z}$. At such point both $\mathcal{Z}'(x)=0$ and $\mathcal{Z}''(x)=0$. By solving these conditions for $x$ and $\lambda$ we obtain
$$\xsfl={\left( 1 - \frac{\gamma }{4} \right) }^
  {\frac{\alpha  + 1}{\gamma }}\xc,\qquad\lsfl^2=\frac{\gamma }{4}\,{\left( \frac{1 - \frac{\gamma }{4}}
      {{\left( \alpha \,{\kappa }^{\alpha  - 1} \right) }^
        {\frac{2}{\alpha  + 1}}} \right) }^
   {\frac{4}{\gamma } - 1},\qquad \epsfl=\frac{3-\beta}{(\alpha-1)(\beta-1)}\frac{\lsfl^2}{\xsfl^2},
   \qquad \gamma<4.$$
$$\det{[\partial^2_{ij}F(\xsfl,\mathcal{Y}(\xsfl))]}=0,
\qquad
\tr{[\partial^2_{ij}F(\xsfl,\mathcal{Y}(\xsfl))]}=\frac{2}{\kappa }\,{\left( \frac{{\left( \alpha \,{\kappa }^{\alpha  - 1} \right) }^
        {\beta  - 1}}{{\left( 1 - \frac{\gamma }{4} \right) }^{\alpha  - 1}} \right)
      }^{\frac{3}{\gamma }}\,\left( \Xi  + \frac{1}{\Xi } \right),
   \quad
  \Xi= \frac{2\,\kappa }{\alpha+1 }\,{\left( \frac{{\left( \alpha \,
           {\kappa }^{\alpha  - 1} \right) }^{5 - \beta }}{{\left( 1 - 
           \frac{\gamma }{4} \right) }^{\alpha  + 3}} \right) }^{\frac{1}{\gamma }}
$$     
This analysis shows 
that the number of shocks for $0<\gamma<4$ is variable, depending on $\lambda$ and is at most three.  The corresponding phase diagrams are shown in Figure \ref{fig:gammaphysunitalphgt} for various $\lambda$ including diagrams for the critical value $\lsfl$ and the purely radial accretion case. For comparison, in Figure \ref{fig:gammaphysunitalph} are shown corresponding phase portraits of the limiting isothermal solution ($\alpha=1$) for $0<\gamma<4$ found in exact form later in section \ref{sec:saisot}.

{\bf D)} {\it Spiralling accretion for $\gamma=0$ ($\alpha>1$, $1<\beta=3-\frac{4}{\alpha+1}<3$)}. 

\begin{figure}[h]
\centering
\includegraphics[width=0.9\textwidth]{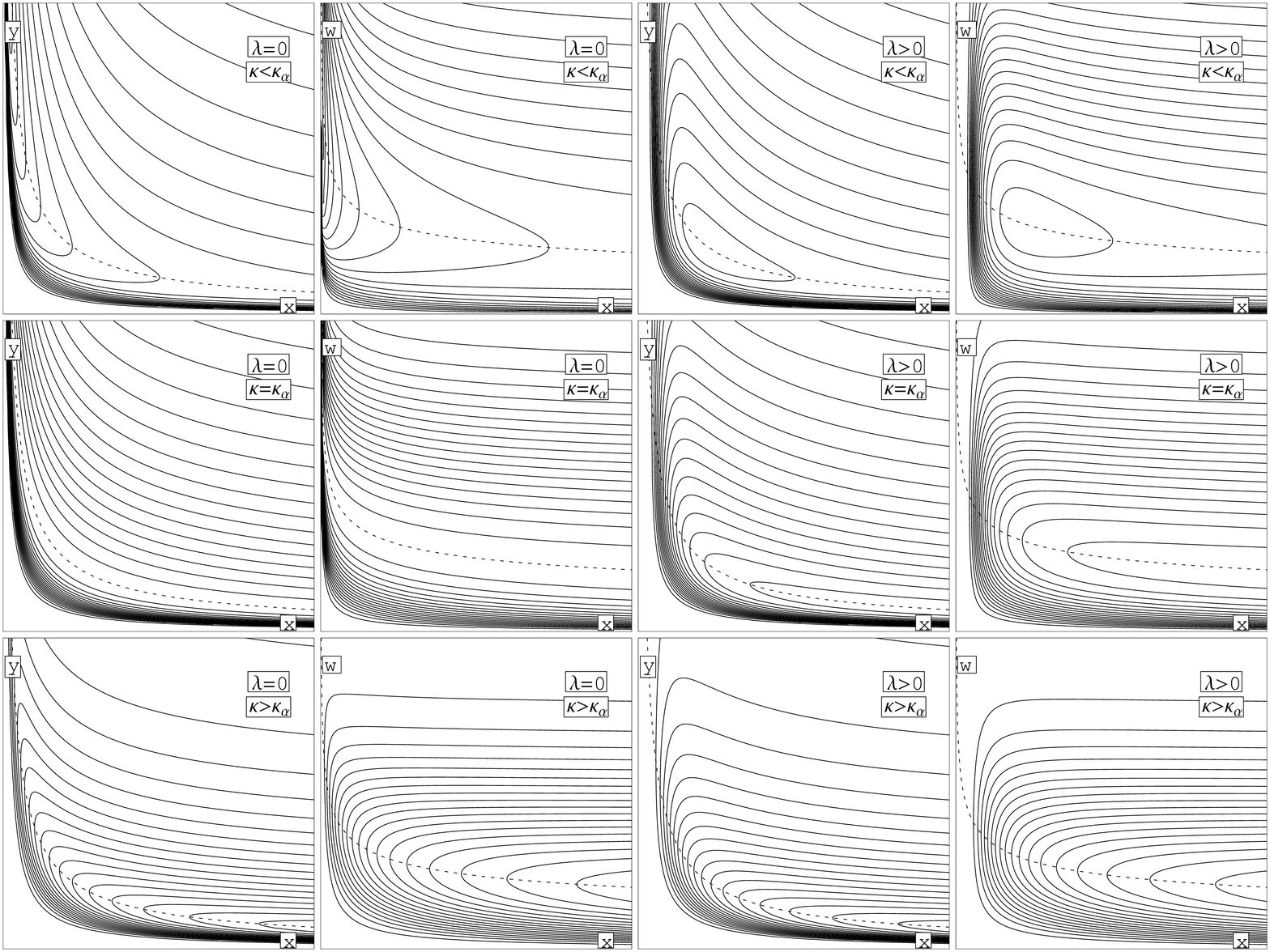}
\caption{\label{fig:gamma0polytr}
$\alpha>1$, $1<\beta=3-\frac{4}{\alpha+1}<3$ ($\gamma=0$) $\diamond$
Radial and spiralling cylindrical accretion of polytropic matter in the power-law potential illustrated with example level lines of the energy surface, shown in the position-density $(x,y)$  and in the position-velocity $(x,w)$ phase planes for various $\kappa$:
$\kappa<\kappa_{\alpha}$ [{\it upper row}], 
 $\kappa=\kappa_{\alpha}$ [{\it middle row}],
$\kappa>\kappa_{\alpha}$ [{\it bottom row}], where $\kappa_{\alpha}=(\!\!\sqrt[\alpha-1]{\alpha})^{{}^{\!-1}}$;
density function for  radial accretion ($\lambda=0$) [{\it 1st column}],
radial velocity component for radial accretion ($\lambda=0$) [{\it 2nd column}],
density function for spiralling accretion ($\lambda>0$) [{\it 3rd column}], and radial velocity component for spiralling accretion ($\lambda>0$) [{\it 4th column}]. }
\end{figure}  

Similarly as it was for radial accretion with $\gamma=0$, we must consider three cases:     

{\it i)} $\alpha\kappa^{\alpha-1}<1$.  
The function $\mathcal{Z}(x)$ starts form $+\infty$ ($\mathcal{Z}(x)\nearrow+\infty$ as $x\searrow0$) and then it is decreasing, attains its $0$ at 
$\xzero$, then it still decreases until it attains at $\xmin$ a global minimum
   $\emin <0$, where 
$$\xzero={\left( \frac{\alpha  - 1}{\alpha  + 1}\,
     \frac{{\lambda }^2}
      {1 - {\left( \alpha \,
            {\kappa }^{\alpha  - 1} \right) }^
         {\frac{2}{\alpha+1 }}} \right) }^
  {\frac{\alpha  + 1}{4}},\qquad
  \xmin=\br{\frac{\alpha+1}{\alpha-1}}^{\frac{\alpha+1}{4}}\xzero,
  \qquad 
  \emin=- \frac{{\left( 1 - 
         {\left( \alpha \,{\kappa }^{\alpha  - 1}
             \right) }^{\frac{2}{\alpha+1 }} \right) 
        }^{\frac{\alpha+1 }{2}}}{\left( \alpha  - 
        1 \right) \,{\lambda }^{\alpha  - 1}}.$$        
        Then the function is increasing and $\mathcal{Z}(x)\nearrow0$ as $x\nearrow+\infty$. As observed in section \ref{sec:qualitative}, the stationary point of the energy surface at the minimum of the shock curve is elliptic. Solutions exist only for $\eps>\emin$. For $\emin<\eps<0$ the solutions are closed loops on the $(x,y)$ plane with shocks at convex $x_a$ and concave $x_b$ turning points, such that $x_0< x_a<\xmin<x_b<+\infty$. For $\eps=0$ there is a limiting
        infinite loop on the right of shock point $x=\xzero$ with the limiting value $y=0$ at $x=+\infty$ for subsonic and supersonic branch. For $\eps>0$ the asymptotics is different: $y\sim \frac{\kappa}{x\sqrt{2\eps}} $ and $w\sim \sqrt{2\eps}$ for a supersonic solution and $y\to y_{\infty}=\br{\frac{\eps(\alpha-1)}{\alpha}}^{\frac{1}{\alpha-1}}$  and $w\sim\frac{\kappa}{x\,y_{\infty}}$ for a subsonic solution, and both solutions terminate at their common convex shock point $0<x_a<\xzero$.  

{\it ii)} $\alpha\kappa^{\alpha-1}=1$. In this case $\mathcal{Z}(x)=\frac{\lambda^2}{2x^2}$ and $\eps>0$. The single shock sonic point at $x_a=\lambda/\sqrt{2\eps}$ is convex. Then the subsonic and supersonic branch of any solution extend from the shock point out to infinity, similarly as it is for radial accretion with $\gamma=0$ and   $\kappa^{\alpha-1}>1$.

{\it iii)} $\alpha\kappa^{\alpha-1}>1$ In this case there is still a single shock sonic point with level lines $x(y)$ convex at that point. In this case the centrifugal term does not change the phase diagram qualitatively compared with the radial accretion. 

The corresponding phase diagrams for all the three cases are shown in Figure \ref{fig:gamma0polytr} including diagrams for the radial accretion.  

\medskip

\noindent
{\bf E)} {\it Spiralling accretion for $\gamma<0$ ($\alpha>1, \ 1<\beta<3-\frac{4}{\alpha+1}$)}

\begin{figure}[h]
\centering
\includegraphics[width=0.9\textwidth]{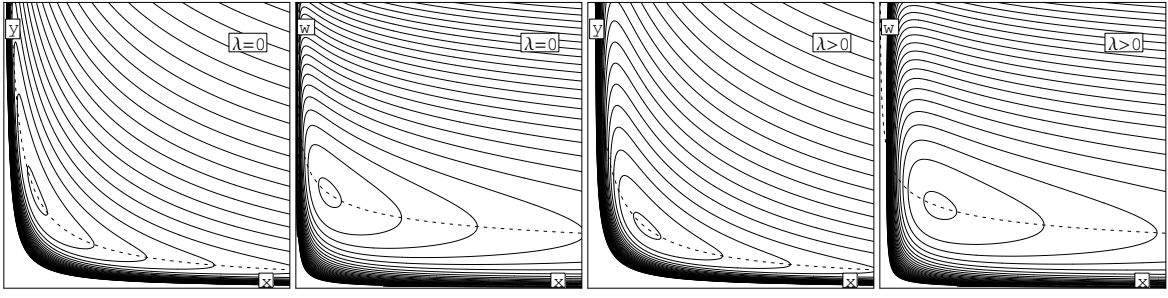}
\caption{\label{fig:gammaneg} $\alpha>1$, $1<\beta<3-\frac{4}{\alpha+1}$\  ($\gamma<0$) $\diamond$
Radial and spiralling cylindrical accretion of polytropic matter in the power-law potential illustrated with example level lines of the energy surface, shown in the position-density $(x,y)$  and in the position-velocity $(x,w)$ phase planes; 
{\it 1st}: density function for radial accretion ($\lambda=0$), 
{\it 2nd}: radial velocity for radial accretion ($\lambda=0$), 
{\it 3rd}: density function for spiralling accretion ($\lambda>0$), 
{\it 4th}: radial velocity for spiralling accretion ($\lambda>0$).
The shock curve is shown with the dashed line.}
\end{figure}

The function $\mathcal{Z}(x)$ 
starts from $+\infty$ ($\mathcal{Z}(x)\nearrow+\infty$ as $x\searrow0$)
and then it is decreasing, becomes negative and attains its global minimum $\eps^{\star}(\lambda)<0$ at some $x=\xc(\lambda)$ (corresponding to the stationary elliptic point of the energy surface) then it monotonically increases until it attains $0$ at spatial infinity. This behaviour is 
qualitatively the same as for radial accretion with $\gamma<0$. Hence, for $\eps^{\star}(\lambda)<\eps<0$  two solutions exist in 
between two shock sonic points, or for $\epsilon>0$, when there is only a single shock sonic point which is a convex turning point, the solutions extend out to spatial infinity. 
Solutions are not possible for $\eps<\eps^{\star}(\lambda)$. The corresponding phase diagrams are shown in Figure \ref{fig:gammaneg} including diagrams for the radial accretion.  

We remind that the asymptotics at the spatial infinity of the positive energy solutions follows from the general results given in equation \eqref{eq:supasympt} and \eqref{eq:subasympt}.

\subsection{\label{sec:saisot} Spiralling accretion for $\alpha=1$ and $\beta>1$}

 The Hamiltonian function in this case reads 
\begin{equation}\label{eq:FBetaAlpha1spiral}F(x,y)\equiv\frac{1}{2}\frac{\kappa^2}{x^2y^2}+\frac{1}{2}\frac{\lambda^2}{x^2}-\frac{1}{\beta-1}\frac{1}{x^{\beta-1}}+\upsilon^2\ln{y}, \qquad x>0,\quad y>0,
\end{equation}
     with $\upsilon$ being the (constant) speed of sound. It has the structure of general equation \eqref{eq:GenAccrEq}.
     On identifying $\mathcal{K}(x)=\frac{1}{2}\frac{\kappa^2} 
{x^2\upsilon^2}$ and $\mathcal{E}(x)=\frac{1}{\upsilon^2}\br{\eps+\frac{1}{\beta-1}\frac{1}{x^{\beta-1}}-\frac{1}{2}\frac{\lambda^2}{x^2}}$  we find that the exact solution in this case reads
$$y(x)=
\exp\sq{\frac{1}{\upsilon^2}\br{\epsilon+\frac{x^{1-\beta}}{\beta-1}-\frac{1}{2}\frac{\lambda^2}{x^2}}+\frac{1}{2} {W}\br{-\frac{\kappa^2}{x^2\upsilon^2}\exp\sq{-\frac{2}{\upsilon^2}\br{\eps+\frac{x^{1-\beta}}{\beta-1}-\frac{1}{2}\frac{\lambda^2}{x^2}}}}},$$ where $W$ is the Lambert function. To obtain a complete level line we must take both 
real branches of function $W$.  The phase diagrams for various parameters $\lambda$ are shown in Figure \ref{fig:gammaphysunitalph} for $\alpha=1$ and can be compared with the respective diagrams for $\alpha>1$ in Figure \ref{fig:gammaphysunitalphgt}.   
     The shock curve corresponding  to the present $F(x,y)$ reads $$y=\mathcal{Y}(x)=\frac{\kappa}{\upsilon x}, \qquad \mathcal{Z}(x)=\frac{1}{2}\,\frac{{\lambda }^2}{x^2} - 
  \frac{1}{\beta  - 1}\,\frac{1}{x^{\beta  - 1}} - 
  {\upsilon }^2\,\log (\frac{\upsilon \,x}
     {\kappa }) + \frac{{\upsilon }^2}{2},\qquad w(x,\mathcal{Y}(x))=\upsilon.$$
       The $y$-dependent part in $F(x,y)$ is bounded from below on the shock curve,  hence 
 the following inequality (analogous to that in equation \eqref{eq:inequ} obtained for $\alpha>1$) is true for solutions:
\begin{equation}\label{eq:ineqx}\frac{{\upsilon }^2}{2} - 
   {\upsilon }^2\,\log (\frac{x\,\upsilon }{\kappa })
   \leqslant \epsilon  - \frac{1}{2}\,
    \frac{{\lambda }^2}{x^2} + 
   \frac{1}{\beta  - 1}\,\frac{1}{x^{\beta  - 1}}\end{equation}
    This means that $\mathcal{Z}(x)\leqslant\eps$ on the level lines $y(x)$ satisfying the equation $F(x,y(x))=\eps$.
The condition is surely satisfied in the limit $x\to+\infty$, hence the solutions may extend out to infinity. 
The asymptotics can be determined perturbatively, as it was done earlier. The branch with 
$y\to0$ and that with finite $y$ as $x\nearrow+\infty$ are following:
 $$y(x)\sim \frac{\kappa}{x}\frac{1}{\sqrt{2\eps +\upsilon^2\ln{\frac{2x^2}{\kappa^2}}}},\qquad y(x)\sim e^{\frac{\epsilon }{{\upsilon }^2}}\br{1+\frac{1}{\br{\beta-1}\upsilon^2}\frac{1}{x^{\beta-1}}}-\frac{1}{2\upsilon^2 x^2}\br{\lambda^2e^{\frac{\eps}{\upsilon^2}}+\kappa^2e^{-\frac{\eps}{\upsilon^2}} }.$$ 
 The second branch can be also obtained by Taylor expansion from the principal branch of exact solution in terms of Lambert's $W$ function.  
 The inequality \eqref{eq:ineqx} is violated in the vicinity of the center for $1<\beta<3$ and $\lambda^2>0$, or for $\beta=3$ and $\lambda^2>1$, and in these cases the solutions cannot reach the centre.    
Similarly as for the power-law potential with $\alpha>1$, the critical value $\beta=3$ (and then also $\lambda^2=1$) is distinguished. The behaviour of function $\mathcal{Z}$ is qualitatively the same to that in Figure \ref{fig:03}. 
The stationary inflection point is possible for function $\mathcal{Z}(x)$ only for $1<\beta<3$ and then the corresponding critical values are
$$\lsfl^2={\upsilon }^2\,\frac{\beta  - 1}{3 - \beta }\,
  {\left( \frac{3 - \beta }{2\,{\upsilon }^2} \right)
      }^{\frac{2}{\beta  - 1}}, \qquad \xsfl={\left( \frac{3 - \beta }{2\,{\upsilon }^2} \right) }^
  {\frac{1}{\beta  - 1}},\qquad \epsfl=\frac{{\upsilon }^2}{\beta  - 1}\,
  \ln\br{\frac{2\,{\upsilon }^2}
     {\left( 3 - \beta  \right)e }\,
    {\left( \frac{\kappa }{\upsilon } \right) }^
     {\beta  - 1}}. $$  For $\lambda^2>\lsfl^2$ there are no extrema and function $\mathcal{Z}(x)$ is monotonically decreasing. Then there is only a single shock point corresponding  to a given energy $\eps$. The point is a convex turning point. For $\lambda^2=\lsfl^2$  the root of $\mathcal{Z}'(x)$ is double and there is only a single shock sonic point for $\eps\neq\epsfl$, while for $\eps=\epsfl$ the subsonic and supersonic solutions corresponding to this energy coalesce as they converge to the shock curve and meet at their end point $(\xsfl,\mathcal{Y}(\xsfl))$ on the phase plane, which is a stationary parabolic point of the energy surface. The two solutions form  in the phase plane a characteristic cusp that can be seen in Figure \ref{fig:gammaphysunitalph}. Such a point should be regarded as a regular sonic point, because the two solutions have a definite slope $y'(x)$ in the limit $x\searrow\xsfl$. 
For $\lambda^2<\lsfl^2$ there are $2$ roots of $\mathcal{Z}'$  
   and hence a minimum followed by a maximum (because $\mathcal{Z}$ is  rapidly decreasing for $x$ small enough).  
   Since the Hessian determinant evaluated on the shock curve and $Z''(x)$ share the same sign, $\det{[\partial^2_{ij}F(x,\mathcal{Y}(x))]}=\frac{2\,{\upsilon }^4}{{\kappa }^2}\,x^2\,\mathcal{Z}''(x)$, the minimum and maximum
   are, respectively, a stationary elliptic point and a stationary hyperbolic point of the energy surface. 
      By comparing with the analogous analysis made in point $C$ of section \ref{sec:spiral}, we see that the structure of level lines as function of $\lambda$ should be similar to that of the power-law potential with $\alpha>1$ and $0<\gamma<4$, with the exception that now $\eps$ is not bound from below. And this expectation is confirmed in Figures   \ref{fig:gammaphysunitalphgt} and \ref{fig:gammaphysunitalph}.
   
\section{Spiralling and radial accretion in the logarithmic potential \ ($\alpha\geqslant1$, $\beta=1$)}
\subsection{Polytropic exponent $\alpha>1$ ($\beta=1$, $\gamma<0$)}
In this section we consider the accretion of polytropic matter onto an infinite homogeneous string. According to equations \eqref{eq:Upotential}
and \eqref{eq:intsurf1} in this case the Hamiltonian function with the centrifugal potential reads
\begin{equation}\label{eq:intsurf2}
F(x,y)=\frac{1}{2}\frac{\kappa^2}{x^2y^2}+\frac{1}{2}\frac{\lambda^2}{x^2}+\ln{x}+\frac{\alpha}{\alpha-1}y^{\alpha-1}, \qquad x>0,\quad y>0.
\end{equation} The corresponding shock curve is 
\begin{equation*}
\mathcal{Z}(x)=\frac{1}{2}\frac{\lambda^2}{x^2}+\lrg{\frac{1}{2}\frac{\alpha+1}{\alpha-1}\br{\alpha{}\kappa^{\alpha-1}}^{\frac{2}{\alpha+1}}x^{-2\frac{\alpha-1}{\alpha+1}}+\ln{x}}
,\qquad x\in(0,\infty)\setminus\br{\sc=\kappa\sqrt[\alpha-1]{\alpha}}.
\end{equation*}
In what follows we investigate qualitatively the simpler case of radial accretion.
For $\lambda=0$ 
equations \reff{eq:hess1} are still valid if we just substitute $\beta=1$. 
Considering that 
now $\gamma=2(1-\alpha)<0$, 
the sign of $\partial^2_{xx}F$ and the sign of the Hessian determinant
at the stationary point are both positive. Therefore, the stationary point of the Hamiltonian is elliptic and a local minimum.
The level lines crossing the sonic shock curve 
are convex for $x<\sc$ and concave for $x>\sc$, as indicated by the sign of $x''(y)$ on both sides of $\xc$:
$$\left.x''(y)\right|_{y=\mathcal{Y}(x)}=\frac{\alpha+1}{\kappa^2}
\pr{\alpha\kappa^{\alpha-1}}^{\frac{3}{\alpha+1}}
\frac{\pr{x/\sc}^{\frac{\alpha+5}{\alpha+1}}}{1-\pr{x/\sc}^
{2\frac{\alpha-1}{\alpha+1}}}.$$
Thus, we obtain  qualitatively the same picture of closed integral curves as for radial accretion in a power-law potential satisfying the same condition $\gamma<0$. The difference is that now there are no solutions extending out to the spatial infinity. The
integration constant $\eps$ is bounded from below by a number  
$$\eps^{\star}=\mathcal{Z}(\sc)=\frac{1+\alpha+2\ln{\alpha}}{2(\alpha-1)}+\ln{\kappa}$$ which may be negative for $\kappa>0$ small enough. Accordingly, the critical accretion rate corresponding to the stationary point for solutions with given specific energy $\epsilon$ is:
\begin{equation}\label{eq:crit_accretion_log}\kappac=\frac{1}{\sqrt[\alpha-1]{\alpha}}\exp{\br{\eps-\frac{\alpha+1}{2(\alpha-1)}}}.\end{equation} The integral curves are now closed for every $\eps>\eps^{\star}$ since $\mathcal{Z}(x)$ diverges to $+\infty$ both when $x\searrow0$ or when $x\nearrow+\infty$, while for $\eps<\eps^{\star}$ there are no solutions at all. It follows from these observations that for $\eps>\eps^{\star}$ there are two solutions - one subsonic and the other supersonic, present between two shock points. The difference between the corresponding $y(x)$ values of these solutions increases with $\eps$. In the limit $\eps\searrow\eps^{\star}$ the 
accretion region reduces to $x=\xc$. By adding the centrifugal term this qualitative result cannot be changed -- for large radii the logarithm term is still dominating in $\mathcal{Z}(x)$, while for small radii the centrifugal term takes over the role of the second power-law potential term in $\mathcal{Z}(x)$.

\begin{figure}[h]
\centering
\includegraphics[width=0.9\textwidth]{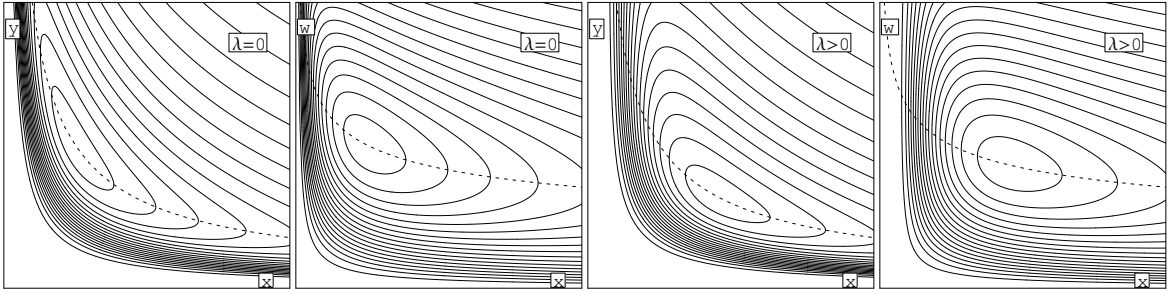}
\caption{\label{eq:logpoly}$\alpha>1$, $\beta=1$ ($\gamma<0$) $\diamond$ 
Radial and spiralling cylindrical accretion of polytropic matter in the logarithmic potential. Example level lines of the energy surface, shown in the position-density $(x,y)$  and in the position-velocity $(x,w)$ planes.
{\it 1st}: density function for radial accretion ($\lambda=0$), 
{\it 2nd}: radial velocity for radial accretion ($\lambda=0$), 
{\it 3rd}: density function for spiralling accretion ($\lambda>0$), 
{\it 4th}: radial velocity for spiralling accretion ($\lambda>0$).
The shock curve is shown with the dashed line. }
\end{figure}

The corresponding phase portraits for radial accretion (as well as for spiralling accretion) are shown in Figure \ref{eq:logpoly}. They can be compared with similar phase portraits for accretion in power-law potential under the same condition $\alpha>1$ and $\gamma<0$ shown in Figure \ref{fig:gammaneg}. 
It is also seen from Figure \ref{eq:logpoly} that the centrifugal term does not change the phase portraits qualitatively. 

\noindent
The problem of the spiralling and radial accretion in the logarithmic potential can be fully solved in analytical way. The
exact solution can again be given in terms of  the Lambert $W$ function. This time, we exchange the meaning of variables $x$ and $y$ in equations \ref{eq:GenAccrEq} and \ref{eq:GenAccrEqSol} and express the solution in the reversed form $x(y)$:
\newcommand{\kyh}{\br{\lambda^2+\frac{\kappa^2}{y^2}}}
\newcommand{\eyh}{\eps-\frac{\alpha}{\alpha-1}y^{\alpha-1}}
$$
x(y)=\exp{\sq{\br{\eyh}+\frac{1}{2}W\br{-\kyh\exp\sq{-2\br{\eyh}}}}},
$$
where we must take both branches of the double valued function $W$ to obtain a complete level line.

\subsection{Radial and spiralling accretion in the logarithmic potential: $\alpha=1$, $\beta=1$ ($\gamma=0$)}

\begin{figure}[h]
\centering
\includegraphics[width=0.9\textwidth]{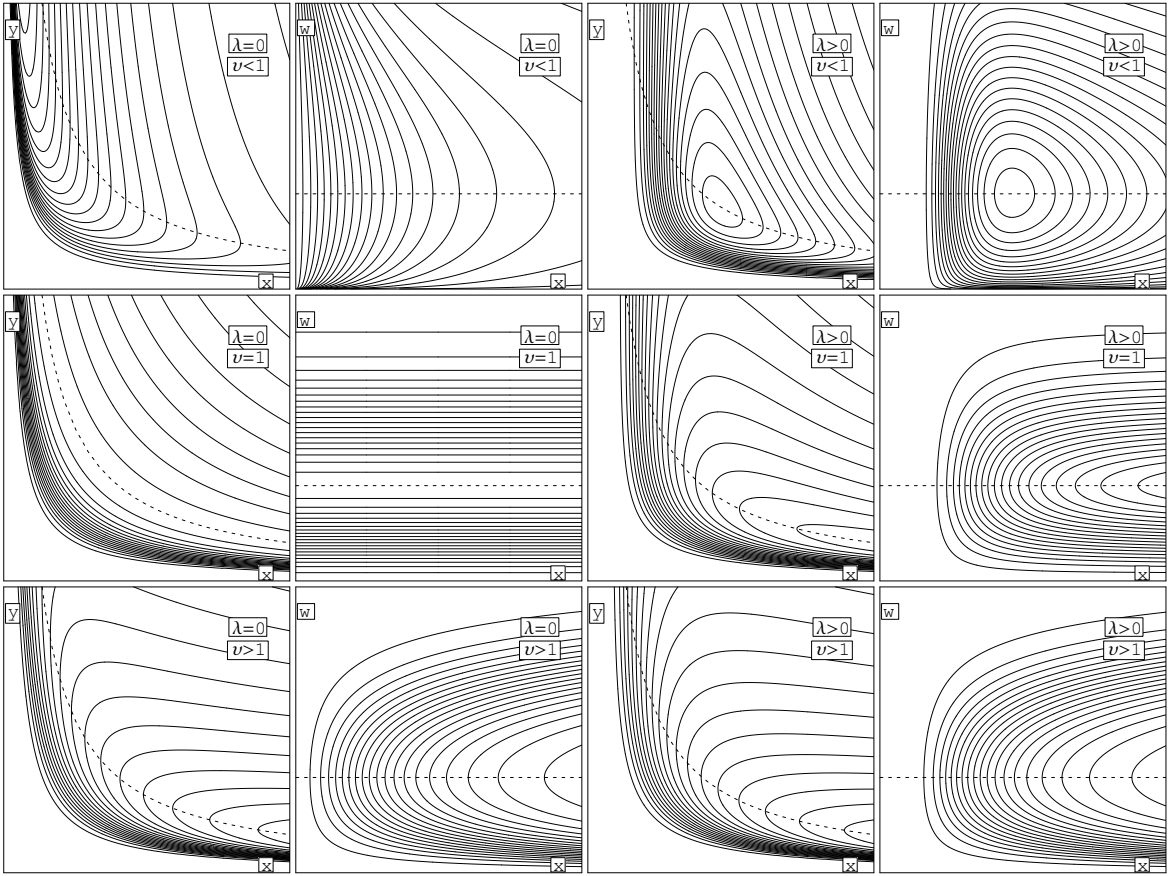}
\caption{\label{fig:loglog}$\alpha=1$, $\beta=1$ ($\gamma=0$) $\diamond$    Radial and spiralling cylindrical accretion of polytropic matter in the logarithmic potential illustrated with example level lines of the energy surface, shown in the position-density $(x,y)$  and in the position-velocity $(x,w)$ phase planes for various $\upsilon$:
 $\upsilon<1$ [{\it upper row}],
 $\upsilon=1$ [{\it middle row}], $\upsilon>1$ [{\it bottom row}]; 
 density function for radial accretion ($\lambda=0$) [{\it 1st column}],
 radial velocity component for radial accretion ($\lambda=0$) [{\it 2nd column}],
density function for spiralling accretion ($\lambda>0$) [{\it 3rd column}], and 
 radial velocity for spiralling accretion ($\lambda>0$) [{\it 4th column}].
The shock curve is shown with the dashed line. }
\end{figure}    
According to equations \eqref{eq:Upotential}
and \eqref{eq:intsurf1}, the Hamiltonian of the spiralling accretion with the logarithmic potential and unit polytropic exponent is
\begin{equation}\label{eq:logF}F(x,y)=\frac{1}{2}\frac{\lambda^2}{x^2}+\frac{1}{2}w^2-\upsilon^2\,\ln{\frac{w}{\kappa}}+(1-\upsilon^2)\ln{x},\qquad w=\frac{\kappa}{x\,y}.\end{equation}
We start with  simpler case of radial accretion ($\lambda=0$).
There are two cases that can be distinguished qualitatively:  $\upsilon=1$ and $\upsilon\neq1$:
\noindent
{\it i).} For $\upsilon=1$, it follows from constant value of the Hamiltonian, that the radial velocity $w$ is also constant (if $\lambda=0$). 
For the minimum value of $\eps$ which we denote here by $\tilde{\eps}$:
\begin{equation}\label{eq:minenerg_log_1}
\tilde{\eps}= \frac{1}{2}+\ln{\kappa},
\end{equation} there is a single regular solution  at $w=1=\upsilon$ (sonic solution): $$y(x)=\frac{\kappa}{x}.$$
 Since $F\to+\infty$ when $w\searrow0$ or when $w\nearrow+\infty$, then for   $\eps=\tilde{\eps}+\delta^2$ there are two solutions 
bifurcating from the previous sonic solution, one subsonic and the other supersonic. For $\delta$ small enough the solutions can be approximated by 
 $$y(x)=\frac{\kappa}{x\,w_{\pm}},\qquad w_{\pm}\approx 1\pm \delta+\frac{1}{6}\delta^2\pm\mathcal{O}(\delta^3).$$ 
   Because $y'(x)=-\frac{\partial_x F}{\partial_y F}=-\frac{y}{x}\ne0$, the accretion is regular with no shocks, either for  sonic accretion  ($\delta=0$) and for supersonic or subsonic accretion ($\delta>0$). 

\noindent   
{\it ii).} For $\upsilon\neq1$, the energy surface has no stationary points. The derivative $y'(x)=-\frac{\partial_x F}{\partial_y F}=-\frac{w^2-1}{w^2-\upsilon^2}\frac{y}{x}$ blows up on the shock curve
$$
%X(x)=x,\qquad 
\mathcal{Y}(x)=\frac{\kappa}{\upsilon\,x}, \qquad \mathcal{Z}(x)=\frac{1}{2}\upsilon^2\ln{\frac{e\kappa^2}{\upsilon^2}}+(1-\upsilon^2)\ln{x}, \qquad (\lambda=0).$$ The image of the shock curve on the $(x,y)$ plane overlaps with the sonic curve (since $w=\upsilon$). 
For a given $\eps$ there is only one single shock sonic point possible, which is located at
\begin{equation}\label{eq:xa_log_1}
x_a=\exp\br{\frac{\eps-\frac{1}{2}\upsilon^2\ln{\frac{e\kappa^2}{\upsilon^2}}}{1-\upsilon^2}}.
\end{equation} The shock point is a turning point of the corresponding level lines.
The turning points are convex for $\upsilon>1$ and concave for $\upsilon<1$, which follows from the sign of  $x''(y)|_{y=\mathcal{Y}(x)}=\frac{2\,\upsilon^4\,x^3}{\kappa^2(\upsilon^2-1)}$ as evaluated on the shock  curve under the condition $x'(y)=0$ at $y=\mathcal{Y}(x)$.
Hence, the accretion is possible for radii $x>x_a$ if $\upsilon>1$ and 
for radii  $x<x_a$ if $\upsilon<1$. These results for radial accretion are summarised in Table \ref{tab:tableA}, while the phase portraits  are shown in Figure \ref{fig:loglog}. In this figure we also show the  corresponding phase portraits for spiralling accretion. As we can see, in this case a second shock point appears for $\upsilon^2<1$ closer to the center. This again can be easily explained. For $\lambda=0$ and $\upsilon^2\neq1$, function $\mathcal{Z}(x)$ behaves as $\ln{x}$ and thus the single shock point for some $\eps$ is concave. If we include the centrifugal term, the resulting $\mathcal Z$ is dominated by this term for small $x$ and is divergent to $+\infty$ at the center, thus it must have a minimum. It is located at $x=\lambda/\sqrt{1-\upsilon^2}$, then we can verify that this point will be an  elliptic point of the Hamiltonian since the Hessian determinant evaluates to a positive number $4\upsilon^4(1-\upsilon^2)/\kappa^2$ at that point. For $\upsilon\geqslant1$ no such a minimum is possible.  
 There is qualitative similarity of the phase portraits to those for  accretion in power-law potential with exponents $\alpha$ and $\beta$ bounded by the same condition $\gamma=0$, see Figure \ref{fig:gamma0polytr}.

Finally, we give the exact form of the solutions valid for any $\lambda$. Based on what has been said in section \ref{sec:alpha1powlaw} for equations of  the general form \eqref{eq:GenAccrEq}, we substitute $y(x)=\exp{\br{\frac{1}{2}\omega(x)+\frac{1}{\upsilon^2}\br{\eps-\ln{x}}}}$ in equation \eqref{eq:logF} and find that $\omega\,\expp{\omega}=-\frac{\kappa^2}{\upsilon^2x^2}\exp\br{-\frac{2}{\upsilon^2}\br{\eps-\ln{x}}}$. Knowing that ${W}(\omega\,\expp{\omega})\equiv\omega$ by the defining property of the Lambert ${W}$ function, we obtain
$$y(x)=
\exp\sq{\frac{1}{\upsilon^2}\br{\epsilon-\ln{x}-\frac{1}{2}\frac{\lambda^2}{x^2}}+\frac{1}{2}{W}\br{-\frac{\kappa^2}{x^2\upsilon^2}\exp\sq{-\frac{2}{\upsilon^2}\br{\eps-\ln{x}-\frac{1}{2}\frac{\lambda^2}{x^2}}}}}.$$
Due to the presence of two logarithms and power functions in equation \eqref{eq:logF} we can similarly obtain a solution in the reversed form of the above solution 
$$x(y)=
\exp\sq{\epsilon-\upsilon^2\ln{y}+\frac{1}{2}{W}\br{-\br{\frac{\kappa^2}{y^2}+\lambda^2}\exp\sq{-2\br{\eps-\upsilon^2\ln{y}}}}}.
$$
In both above forms of the solution we must take both real branches of the double valued function $W$ in order to obtain a complete level line.

   \section{Conclusions}
   
We investigated cylindrically symmetric accretion of inviscid polytropic matter in the power-law and logarithmic potentials both for zero angular momentum (purely radial flow) and non-zero angular momentum (spiralling flow). As we have seen, this model belongs to a class of fully cylindrically symmetric hydrodynamical models of horizontal flows  onto the  
infinite symmetry axis which, as idealisations of physical situations such as spiral arms or gas filaments, have been considered in various contexts in the astrophysical literature, including, in particular, models of purely radial flow of infinite cylinders (e.g. radial collapse of self-gravitating isothermal cylinders) or even equilibrium solutions such as infinite gravitating polytropic cylinders \citep{1964ApJ...140.1056O}.
We analysed the character of the flow
solutions in the steady-state and ignoring self-gravity. 
The analysis was straightforward as it was tantamount to studying the isocontours of the Hamiltonian of an equivalent dynamical system with one degree of freedom. We investigated also the asymptotics of solutions. 
Although so simple, the model of spiralling accretion turned out to be characterized by a very rich family of phase diagrams of the flow, which in some cases show qualitative similarities to the classical Bondi accretion model, which we also studied here as an example with the mathematical tools we applied to spiralling accretion.
Similarly to the isothermal Bondi model, the isothermal radial accretion solutions could be expressed in terms of the Lambert $W$ function. This was also possible for isothermal spiralling flow solutions. Moreover, we found exact radial and spiralling flow solutions in the logarithmic potential for arbitrary polytropic exponent. We presented also two simple exact radial and spiralling flow solutions expressed in terms of elementary functions and which have non-trivial phase diagrams (equations \eqref{eq:exactsoldelta} and \eqref{eq:solx}). 

\smallskip

{\it Radial accretion solutions.} Within the range $1\leqslant\alpha\leqslant\frac{5}{3}$ of the polytropic exponent considered physical in the classical Bondi problem, the purely radial cylindrical accretion differs qualitatively from the Bondi accretion. In particular, with these $\alpha$ values,  the spherical accretion solutions may extend to all of space with suitable accretion rate, whereas, for the counterpart cylindrical accretion solutions in the logarithmic potential ($\beta=1$) the flow is spatially bounded  between two shock sonic points.
Thus, for $\beta=1$ and $\alpha>1$ there are no global solutions extending to all of space. Global solutions are possible in the logarithmic potential  only for $\alpha=1$ (by which the equation of state is distinguished),  provided that the speed of sound $c$ precisely equals the velocity scale  defined by the logarithmic potential, \ie $c=\sqrt{G\,\mu}$ (with $\mu$ being the amount of mass per unit length). Then, for the accretion rate not higher (or the specific energy not lower) than some critical value, the solutions may extend to all of space. 
For $\beta>1$,  there is a number of qualitatively distinct radial accretion solutions possible, depending on 
the model parameters -- there can be both solutions with phase diagrams resembling qualitatively the Bondi accretion or cylindrical radial accretion in the logarithmic potential. Formally, if we also allow for the classical Bondi accretion with 
$\alpha$ higher than the critical value $\alpha=5/3$, then we observe a correspondence between spherical and cylindrical purely radial accretion solutions. Namely, there are three kinds of qualitatively different phase diagrams possible in each case 
depending on the signature of the Hessian matrix evaluated at the stationary point of the Hamiltonian. For negative specific energy, $\epsilon<0$, solutions with the stationary elliptic point on the phase diagram are spatially bounded  between two shock sonic points, while for $\epsilon>0$ there is a single shock sonic point and the solution extends from that point out to spatial infinity. In the case of classical Bondi accretion, the phase diagram has a stationary hyperbolic, parabolic or elliptic point for $\alpha<5/3$, $\alpha=5/3$ and  $\alpha>5/3$, respectively, while  the three types of stationary points occur respectively for $\gamma>0$, $\gamma=0$ and $\gamma<0$ in the case of cylindrical radial accretion, where $\gamma\equiv 4+(\alpha+1)(\beta-3)$. In cylindrical symmetry one can choose $\beta$ such that solutions with $\alpha>5/3$ and with $\alpha<5/3$ are possible, thus this value of polytropic exponent is not critical. 
 
Of those solutions possible in the investigated radial accretion model, as physical should be considered both solutions without shock points
(corresponding to sub-critical accretion rates) as well as solutions with  regular sonic points (without shocks) corresponding to the critical accretion rate. Solutions with shocks are commonly regarded as unphysical. The description of shocks, likely to involve energy dissipation processes, goes beyond the scope of the simple accretion model we are considering. In  this model shock points are boundary or turning points of solutions. Such solutions do not provide a spatially global description of the accretion process. 

Spatially global radial accretion solutions with $\beta>1$ are possible for properly chosen accretion rates, when the parameters 
$\alpha$ and $\beta$ satisfy the inequality $\gamma\geqslant0$ (at some positive specific energy, $\epsilon>0$). For $\gamma>0$, at the critical accretion rate there is a single regular sonic point (\ie with no shock), and when the accretion rate is higher than the critical one, the solution is bounded in between two shock sonic points and the solution loses its global character. 
For radial accretion with $\gamma=0$, we obtain a cylindrical counterpart of Bondi accretion with the critical polytropic exponent $\alpha=5/3$ and a single shock sonic point. In the limit of the critical accretion rate, the shock moves to the center, and so we obtain a physical spatially global  solution. When
$\gamma<0$ then (at positive specific energy, $\epsilon>0$) there is always a single shock sonic point present that cannot be displaced to the center, as so the accretion with such parameters should be considered as unphysical. For $\gamma<0$ (at negative specific energy, $\epsilon<0$) there are two shock sonic points and solutions may exist only in between the two shock sonic points and thus are also non-physical. 

\smallskip

{\it Spiralling accretion solutions.} The class of investigated spiralling flow solutions (with non-zero angular momentum) seems astrophysically more relevant as models of accretion disks. We think that the status of shock points for the spiralling accretion solutions is less clear than for the radial accretion solutions, because the presence of shock points is generic for $1<\beta<3$. For non-zero angular momentum, spatially global solutions without shocks are possible for $\beta>3$ (examples of such solutions can be seen in the last two columns of Figures \ref{fig:gammabeyondfour} and \ref{fig:gammabeyondfourisot}). In the case when spatially global solutions are possible,
there is a critical accretion rate (analogous to that for Bondi accretion), 
below which solutions are global and above which solutions are not global and bounded by shock sonic points. At the critical rate the solution is  global and passes through a regular sonic point. 
This qualitative observation easily follows from inequality \eqref{eq:inequ} in the case of power-law potentials, and similarly this can be seen from inequality \eqref{eq:ineqx} in the case of logarithmic potential.
The centrifugal potential deforms the phase diagrams close to the centre to such extent that 
solutions gain new features not observed for radial solutions, like solutions with three shock sonic points or even solutions which rapidly end in a quasi-shock at some radius where the density derivative or the radial velocity derivative still remain finite, as can be seen in the third column of Figures \ref{fig:gammaphysunitalphgt} and \ref{fig:gammaphysunitalph}.  
Frequent occurrence of shocks in the accretion with non-zero angular momentum is physically clear due to the repulsive effect of the centrifugal potential divergent as $x^{-2}$ 
preventing matter from reaching the centre and dominates the attractive effect of the power-law potential with $\beta<3$.

The question arises how to interpret the frequent occurrence of shocks for spiralling accretion?  The convex turning point (or shock point) of an external solution on the branch with vanishing velocity at infinity  (like those in the bottom row of Figure \ref{fig:loglog} or similar) could be interpreted as the internal cylindrical boundary of the accretion 'disk'. Solutions of this type are double-valued. Similarly double-valued solutions in Bondi model, called bouncing solutions \citep{Pett1980},  are usually excluded from consideration. However, solutions of this type may represent parts of a correct solution if shocks are present, as was pointed out in the context of Bondi accretion \citep{FKR2002}. 
For spiralling accretion, the physical interpretation would be such that the polytropic equation of state offers a viable approximation for radii only above the shock point, while the shock point indicates the limit of applicability of the model with polytropic equation of state. In a more realistic model  one should expect in this region certain energy dissipation processes to take place. As so, below the shock point we should use a different equation of state, the more that outside the 'disk' we have to do with a different physical situation. 

As a remedy for the problem of shocks in the present  model with the polytropic equation of state, we suggest to consider an improved  model with viscous terms. The terms involve spatial derivatives of the velocity components (which diverge at shocks in the present model). Then we should expect that in regions where the viscous correction from the derivatives is not important, the present model would work well and not much differ in its predictions from the improved model. Similarly, in regions where  the derivatives diverge in the present model, the viscous terms should modify solutions to such extent that the singularities do not appear. 
As to interior solutions with external boundary shocks (concave turning points), we leave open the question of possible physical interpretation. However, for solutions with the same type of points in Bondi model \cite{Pett1980} suggested
to consider discontinuous flows which match at the shock point with the external 
globally sonic Bondi solution. This idea could be also applied to the cylindrical accretion in the case of diagrams with hyperbolic stationary point, in particular, for solutions with concave turning point whose radial velocity is zero on the symmetry axis (this is the case, for $0<\gamma\leq4$ and lambda small enough   or for $\gamma>4$ and any $\lambda$).

\medskip

\noindent
{\bf Acknowledgements}.
We gratefully acknowledge the anonymous referee for many
constructive advices and suggestions which have contributed
to improve this paper.

\appendix

\section{Spiralling accretion model from general accretion disks}
In this section we show how the spiralling accretion model can be arrived at by considering accretion disks in spherical potentials.
We start with considering a general axi- and plane-symmetric accretion disk in the field of a spherical potential. We may solve the partial differential equations perturbatively in many ways.    As we consider an accretion disk we may assume it being entirely contained between two planes parallel to the equatorial mid-plane ($z=0$) of the reflection symmetry. One example of perturbations would be by application of Fourier series in $z$ with some periodic boundary conditions on the planes. However, for the purpose of simple illustration of how the model can be put in a broader context, we may also proceed  with even easier an approach.
In this region bounded by parallel planes we may confine ourselves to considering solutions analytic in $z$ in the form of infinite power series $f(R,z)\sim\sum_k \frac{z^k}{k!}\tcff{f}{k}$ 
with expansion coefficients $\tcff{f}{k}$ defined on the mid-plane $z=0$ as functions of $R$. The plane-symmetry means that the density and the horizontal velocity components are series with even powers of $z$ while the vertical velocity component is a series with odd powers of $z$. Working in cylindrical coordinates $R,\phi,z$ to the linear accuracy in $z$ both for a spherically-symmetric and a cylindrically symmetric potential, we obtain 
following $\sigma$-family of flow models with constant specific angular momentum parameter $J$:
\newlength{\blabla}
 \settowidth{\blabla}{{\tiny{in spherical}}}
 $$
\begin{array}{ccc}
\frac{J^2}{R^3} - \tcff{v_R}{0}\,\tcff{v_R}{0}' - {U}' - \dot{\Pi}[\tcff{\rho}{0}]\frac{\tcff{\rho}{0}'}{\tcff{\rho}{0}} = 0,
&\tcff{v_{\phi}}{0}=\frac{J}{R},
&-\frac{\dot{\Pi}[\tcff{\rho}{0}]}{\tcff{\rho}{0}}\tcff{\rho}{2}= { \sigma^2  + 
    \tcff{v_R}{0}\,{\sigma}'} +\left\{\begin{array}{cl}R^{-1}\,{U}'&
    ${\parbox{\blabla}{\tiny in spherical field}}$\\
    0& 
    ${\parbox{\blabla}{\tiny in cylindrical field}}$
    \end{array}\right.,
\\
{\br{R\,\tcff{\rho}{0}\,\tcff{v_R}{0}}}'=-R\,\tcff{\rho}{0}\,\sigma,
&\tcff{p}{0}=\Pi[\tcff{\rho}{0}].
&
    \end{array}
$$
  The $1$st, $2$nd and $3$rd (upper row) equations are implied by the radial, azimuthal and vertical components of the Euler hydrodynamical equations, the $4$th and $5$th (bottom row) equations follow from the continuity of flow equation and the equation of state. All functions in above equations are defined on the equatorial plane $z=0$ and are dependent on $R$ only. The prime sign denotes differentiation with respect to $R$, while $\dot{\Pi}(\rho)\equiv \frac{\mathrm{d}\Pi(\rho)}{\mathrm{d}\rho}$. 
Here, $\sigma\equiv\tcff{v_z}{1}$ is arbitrary function specifying the behaviour of the vertical velocity component to the first order in $z$: $v_z(R,z)\approx z\,\sigma(R)$.  Having chosen a particular $\sigma(R)$, 
  we can solve the first column equations for $\tcff{v_R}{0}$ and $\tcff{\rho}{0}$ and hence obtain $v_R(R,z)$ and $\rho(R,z)$ to the first order in $z$; then the $3$rd equation implies the pressure off the equatorial plane: $p(R,z)\approx \Pi[\tcff{\rho}{0}(R)]+\frac{1}{2}\tcff{\rho}{2}(R)z^2$ required to support the flow in this approximation. Thus, in this approximation the equations governing the horizontal flow in the spherical and cylindrical potential differ only by the correction to the pressure off the equatorial plane which is set uniquely by the $3$rd equation once the other equations have been solved.   
  This means, that to the linear order, we may discard the $3$rd equation as having no effect upon the flow close to the equatorial plane.
Moreover, as follows from the $4$th equation, the contribution from the vertical flow component of the accretion disk can be omitted when for any $R$ inside the accretion disk of some outer radius  $R_{\mathrm{D}}$ and for a given accretion rate $A$: $$\left|\int\limits_{R}^{R_{\mathrm{D}}}r \rho(r)\sigma(r)\ud{r}\right|\ll A,$$ then also $R\rho(R)v_R(R)\approx A$.
In this case the solutions are well approximated by a simpler model of horizontal spiralling accretion (with constant accretion rate $A$) defined by the reduced system of equations
\begin{equation}\label{eq:A1}\frac{{{v_{\phi }}^2}}{R} - {v_R}\,{v_R}' - U' - \frac{p'}{\rho} = 0,\quad
v_{\phi}=\frac{J}{R},\quad 
v_R=-\frac{A}{R\,\rho}.\end{equation}
The equations are the same as for horizontal flow of test matter under full cylindrical symmetry. As we have just seen, the simple equations also describe 
general cylindrically-symmetric flow to the leading order (provided the vertical components of flow is small in the sense of the above integral) and to first approximation they also describe accretion disks in the spherical potential. 

The above perturbation scheme could be continued at least in principle.
For every even approximation (of order $2k$) we would obtain $3$ new ordinary differential equations for Taylor coefficient functions $\tcff{v_R}{2k}$, $\tcff{v_{\phi}}{2k}$, $\tcff{v_z}{2k+1}$, while for every odd approximation (of order $2k+1$) we would obtain a single algebraic expression for $\tcff{\rho}{2k}$. Equations of a given order depend on solutions of all lower orders. In consequence of this, the perturbative solution in a given potential depends on single arbitrary function $\sigma$ that specifies the boundary condition on the  equatorial plane.

\bibliography{accretion}
\bibliographystyle{apj}

\end{document}